\documentclass[11pt,a4paper]{article}
\usepackage{jheppub}
\usepackage{bbm}
\usepackage{pdfpages}
\usepackage{epstopdf}
 \usepackage{graphicx}

\usepackage{titlesec}

\titleformat{\paragraph}
{\normalfont\normalsize\bfseries}{\theparagraph}{1em}{}
\titlespacing*{\paragraph}
{0pt}{3.25ex plus 1ex minus .2ex}{1.5ex plus .2ex}

\input pix.sty

\newcommand{\sumint}[1]{\mbox{$\sum$}\!\!\!\!\!\!\!\int_{#1}}

\renewcommand{\nr}[1]{(\ref{#1})}

\renewcommand{\fig}{fig.~}

\newcommand{\dd}{\mathrm{d}}
\newcommand{\tinymsbar}{{\overline{\mbox{\tiny\rm{MS}}}}}

\newcommand{\Nc}{N_{\rm c}}

\newcommand{\gB}{g_\rmii{B}}
\newcommand{\mE}{m_\rmii{E}}

\newcommand{\rmO}{{\mathcal{O}}}
\newcommand{\bmu}{\bar\Lambda} %{\bar\mu}

\def\lsi{\raise0.3ex\hbox{$<$\kern-0.75em\raise-1.1ex\hbox{$\sim$}}}
\def\gsi{\raise0.3ex\hbox{$>$\kern-0.75em\raise-1.1ex\hbox{$\sim$}}}

\newcommand{\rmii}[1]{{\mbox{\tiny\rm{#1}}}}

\newcommand{\im}{\mathop{\mbox{Im}}}

\newcommand{\Tint}[1]{{\hbox{$\sum$}\!\!\!\!\!\!\!\int\,}_{\!\!\!\!\raise-0.9ex\hbox{$\scriptstyle{#1}$}}}
\newcommand{\Tinti}[1]{{{\Sigma}\!\!\!\!\raise0.3ex\hbox{$\int$}_\rmii{${#1}$}}}
\renewcommand{\Tint}[1]{\sumint{#1}}

\newcommand{\bi}{\begin{itemize}}
\newcommand{\ei}{\end{itemize}}
\newcommand{\hide}[1]{ }

\newcommand{\Jt}[2]{\mathcal{J}_\rmii{#1}^\rmii{#2}}
\newcommand{\It}[2]{\mathcal{I}_\rmii{#1}^\rmii{#2}}

\allowdisplaybreaks

\title{On the infrared behavior of the shear spectral function in hot Yang-Mills theory}

\author[1]{Aleksi Vuorinen}
\author[2]{and Yan Zhu}

\affiliation[1]{Department of Physics and Helsinki Institute of Physics, P.O.Box 64, FI-00014 University of Helsinki, Finland}
\affiliation[2]{Departamento de F\'{i}sica de Part\'{i}culas and
IGFAE, Universidade de Santiago de Compostela, E-15706 Santiago de Compostela, Galicia, Spain}

\emailAdd{aleksi.vuorinen@helsinki.fi}
\emailAdd{yan.zhu@usc.es}

\abstract{We revisit the determination of the two-loop spectral function in the shear channel of hot Yang-Mills theory. Correcting a technical error in an earlier computation and extending the result with a leading order Hard Thermal Loop resummation is seen to improve the infrared behavior of the quantity significantly. This makes it possible to straightforwardly use the result in the corresponding imaginary time correlator and the shear sum rule.}


\begin{document}

\preprint{HIP-2014-02/TH}

\maketitle

%%%%%%%%%%%%%%%%%%%%%%%%%%%%%%%%%%%%%%%%%%%%%%%%%%%%%
\section{Introduction}\label{intro}
%%%%%%%%%%%%%%%%%%%%%%%%%%%%%%%%%%%%%%%%%%%%%%%%%%%%%

The shear viscosity of the quark gluon plasma (QGP) has been identified as one of the key parameters describing the medium, having particular impact on the hydrodynamic behavior of the matter produced in heavy ion collisions at RHIC and at the LHC (see e.g.~\cite{Tannenbaum,Muller,Romatschke:2009im,Shen:2011zc,Brambilla:2014jmp}). Despite the strong experimental and phenomenological motivation, a nonperturbative first principles tool to predict its value is, however, still lacking,  even though an extensive amount of work has been devoted to the topic in the weak coupling \cite{Aarts:2002cc,Arnold:2003zc}, lattice \cite{Meyer:2007ic,Meyer:2007dy,Meyer:2011gj} and gauge/gravity frontiers \cite{Kovtun:2004de,Kovtun:2011np,Rebhan:2011vd}. The main issue preventing a straightforward lattice determination of the parameter is its inherently Minkowskian nature: According to the Kubo formulae, the viscosity is available as the zero frequency limit of the corresponding spectral function,
\ba
\eta &=& \lim_{\omega\to 0} \fr{\rho_\eta(\omega)}{\omega}\, ,
\ea
obtained from the imaginary part of a retarded (Minkowskian) Green's function.

One promising attempt to overcome the difficulties involved in the determination of transport coefficients is the analytic continuation of lattice results for Euclidean imaginary time correlators,
\begin{equation}
 G(\tau) =
 \int_0^\infty
 \frac{{\rm d}\omega}{\pi} \rho(\omega)
 \frac{\cosh\Big[\! \left(\frac{\beta}{2} - \tau\right)\omega\Big]}
 {\sinh\frac{\beta \omega}{2}}\, ,\quad \quad 0<\tau <\beta \, ,  \la{int_rel}
\end{equation}
proposed and tested in \cite{Burnier:2011jq,Burnier:2012ts,Burnier:2012ze}. An integral part of this program is the analytic subtraction of short-distance divergences from the results, i.e.~obtaining as much analytic information of the ultraviolet (UV) structure of the spectral function as possible. This task, which should be carried out separately at zero and finite temperature, is most conveniently addressed using the machinery of perturbation theory.

The perturbative evaluation of thermal spectral functions becomes a complicated task beyond leading order, and it is only rather recently that progress in this direction has been achieved. For pure SU($N$) Yang-Mills theory, defined via the Euclidean action (note that we work in Euclidean metric throughout the paper)
\ba
 S_\mathrm{E} &=& \int_{0}^{\beta} \! \dd \tau \int \! {\rm d}^{D-1}\vec{x}
 \, \frac{1}{4} F^a_{\mu\nu} F^a_{\mu\nu} \, \equiv \, \int_x \, \frac{1}{4} F^a_{\mu\nu} F^a_{\mu\nu} \,, \label{action}
\ea
with $D\equiv 4-2\epsilon$, $\beta\equiv 1/T$ and
\ba
F^a_{\mu\nu} &\equiv&  \partial_\mu A^a_\nu - \partial_\nu A^a_\mu + \gB f^{abc} A^b_\mu A^c_\nu\, ,
\ea
a next-to-leading order (NLO) spectral function was first determined in the bulk channel in \cite{Laine:2011xm}, building on the earlier work of \cite{Burnier:2008ia,Laine:2010tc,Laine:2010fe} (see also \cite{Moore:2008ws}). Since then, the techniques developed in these papers have been further generalized to include the case of non-vanishing external three-momenta \cite{Laine:2013vpa}, motivated by applications beyond QCD, most importantly studies of the leptogenesis scenario.

In the context of QGP physics, an obvious goal is to extend the bulk calculations to the technically significantly more tedious shear channel. This challenge has indeed been addressed first on the level of the Operator Product Expansion (OPE) in \cite{Schroder:2011ht}, and later by considering the full NLO spectral function in \cite{Zhu:2012be} (see also the related $T=0$ work of \cite{Zoller:2012qv,Zoller:2014dca}). While consistent with known sum rules \cite{Romatschke:2009ng,Meyer:2010gu} as well as the arguments of \cite{CaronHuot:2009ns} concerning the UV behavior of various Green's functions, the latter of these calculations delivered somewhat surprising results. Most importantly, it was observed that in the small-frequency limit, the perturbative spectral function tends to a constant. While not directly alarming --- after all, it is known that to reach the true $\omega \to 0$ limit of the quantity, one needs to perform an elaborate resummation \cite{Arnold:2003zc} --- this implied that it was not possible to straightforwardly apply the result to the determination of the imaginary time correlator of eq.~(\ref{int_rel}).

The situation described above clearly calls for a more detailed study of the infrared (IR) behavior of the shear spectral function, utilizing a Hard Thermal Loop (HTL) resummation to extend the region of validity of the results of \cite{Zhu:2012be} to frequencies of order $gT$. This is not only important for the sake of academic interest, but also to aid the eventual extraction of the shear viscosity from Euclidean lattice data and to facilitate more accurate comparisons with recent lattice and AdS/CFT calculations \cite{Huebner:2008as,Iqbal:2009xz,Springer:2010mf,Springer:2010mw,Kajantie:2010nx,Kajantie:2011nx,Kajantie:2013gab}. In addition to the HTL exercise, we will, however, also revisit the unresummed calculation of \cite{Zhu:2012be}, performing a completely independent evaluation of the shear spectral function to test the correctness of our earlier results.

As the setup of our unresummed calculation is in practice identical to that of \cite{Zhu:2012be}, we refrain from presenting a lengthy introduction to the technical machinery involved in this part of the work. Instead, we will simply walk the reader through the necessary notations and definitions in section 2. After this, we explain the details of the HTL resummation in section 3, and subsequently present and analyze our results in section 4. Section 5 is finally devoted to drawing conclusions, while appendices A--C contain some lengthy definitions and technical details concerning the evaluation of the master sum-integrals encountered.

%%%%%%%%%%%%%%%%%%%%%%%%%%%%%%%%%%%%%%%%%%%%%%%%%%%%%%%%%
\section{Unresummed calculation}\label{unresummed}
%%%%%%%%%%%%%%%%%%%%%%%%%%%%%%%%%%%%%%%%%%%%%%%%%%%%%%

We are interested in connected Green's functions of specific components of the energy-momentum tensor of SU($N$) Yang-Mills theory
\ba\la{eq:T}
 T_{\mu\nu} &=& \frac{1}{4} \delta_{\mu\nu} F^a_{\alpha\beta} F^a_{\alpha\beta} -F^a_{\mu\alpha} F^a_{\nu\alpha} \, ,
\ea
denoted by
\ba
 G_{\mu\nu,\alpha\beta}(x)  &\equiv& \langle T_{\mu\nu}(x)\: T_{\alpha\beta}(0)\rangle_c\, ,
\label{eq:corr_def_coord}
\ea
and in particular in the associated (momentum space) spectral functions,
\ba
 \rho_{\mu\nu,\alpha\beta} (\omega) 
 &\equiv& \im \Bigl[ \widetilde{G}_{\mu\nu,\alpha\beta}(P) 
 \Bigr]_{P \to (-i[\omega + i 0^+],\vec{0})}
 \,. \label{rho_general}
\ea
As discussed already in \cite{Schroder:2011ht}, the most convenient way to access the momentum space shear correlator $\widetilde G_{12,12}(P)$ in dimensional regularization proceeds by introducing the projection operator
\ba
X_{\mu\nu,\alpha\beta}(P) &\equiv& P_{\mu\nu}^{T}P_{\alpha\beta}^{T}
	-\frac{D-2}{2}(P_{\mu\alpha}^{T}P_{\nu\beta}^{T}+P_{\mu\beta}^{T}P_{\nu\alpha}^{T}) \,,\label{eq:xproj}
\ea
where $P_{\mu\nu}^{T}(P)$ is a usual transverse projector orthogonal to the four-vectors $P$ and $U\equiv(1,\mathbf{0})$. Choosing the spatial momentum $\mathbf{p}$ to point in the $x_{D-1}$ direction, this gives
\begin{equation}
    X_{\mu\nu,\alpha\beta}\, \widetilde G_{\mu\nu,\alpha\beta}(P) = -D(D-2)(D-3)\, \widetilde G_{12,12}(P) \,,
\end{equation}
which prompts the definition
\ba
 \widetilde G_\eta(P) \equiv 2 X_{\mu\nu,\alpha\beta} \, \widetilde G_{\mu\nu,\alpha\beta}(P) \label{eq:g_eta_p}
\ea
and leads to the $\epsilon = 0$ identity
\ba
\rho_\eta(\omega)&=&-16\,\rho_{12,12}(\omega)\, . \label{sheardef}
\ea

%%%%%%%%%%%%%%%%%%%%%%%%%%%%%%%%%%%%%%%%%%%%
\begin{figure}
\centering
\epsfxsize=10cm \epsfbox{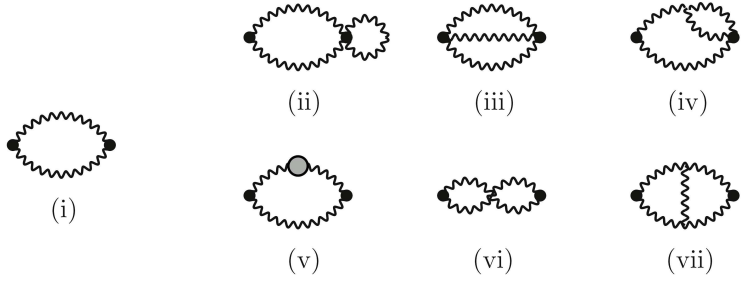}
\caption{The one- and two-loop Feynman diagrams contributing to the NLO shear spectral function in SU($N$) Yang-Mills theory. The curly line corresponds to the gluon field, while the grey blob in (v) denotes the one-loop gluon self energy.}
\label{figs}
\end{figure}
%%%%%%%%%%%%%%%%%%%%%%%%%%%%%%%%%%%%%%%%%%%%

As explained in \cite{Zhu:2012be}, to NLO the shear spectral function obtains contributions from the 1- and 2-loop graphs of \fig\ref{figs}, and can thus be written in the form ($d_A\equiv \Nc^2-1$)
\be
\frac{\rho_\eta (\omega)}{4 d_A \Lambda^{2\epsilon}} =
\rho^{ }_\text{(i)} (\omega) + \gB^2 \Nc\big\{\rho^{ }_\text{(ii)} (\omega)+ \rho^{ }_\text{(iii)} (\omega)+ \rho^{ }_\text{(iv)} (\omega)+ \rho^{ }_\text{(v)} (\omega)+ \rho^{ }_\text{(vi)} (\omega)+ \rho^{ }_\text{(vii)} (\omega)\big\}\, , \label{rhoeta1} 
 %%%%%%%%%%%%%%%
\ee
where each term corresponds to the graph with the same index. In terms of the master integrals defined in appendix \ref{masters}, these functions read (in an arbitrary covariant gauge)
\ba
\rho^{ }_\text{(i)} (\omega) &\equiv& -\fr{D(D-2)(D-3)}{8}\rho^{ }_{\Jt{b}{0}} (\omega) -D(D-3)\rho^{ }_{\Jt{b}{1}} (\omega)
- (D-2)(D-3)\rho^{ }_{\Jt{b}{2}} (\omega)\, , \la{rhoLO}
\\ %%%%%%%%%%%%%%%
\rho^{ }_\text{(ii)} (\omega) &\equiv& \frac{1}{2} D (D-2)^2 (D-3) \rho^{ }_{\It{b}{0}}+2D(D-2) (D-3) \rho^{ }_{\It{b}{2}} \,, \la{rhoI}
\\ %%%%%%%%%%%%%%%
\rho^{ }_\text{(iii)} (\omega) &\equiv& \frac{1}{6} D(D-2) (D-3) \rho^{ }_{\It{f}{0}}-\frac{3}{2} D(D-3) (D-4) \rho^{ }_{\It{f}{1}} \,, \la{rhoC}
\\ %%%%%%%%%%%%%%%
\rho^{ }_\text{(iv)} (\omega) &\equiv&- D (D-2) (D-3) \rho^{ }_{\It{b}{0}}+D (D-3) (2 D-5) \rho^{ }_{\It{b}{1}}-3D\left(D-3\right) \rho^{ }_{\It{b}{2}}
\nn&-&\frac{19}{12}D(D-2)(D-3) \rho^{ }_{\It{f}{0}}+\frac{1}{4}D (D-3)(2D-25) \rho^{ }_{\It{f}{1}}+\frac{3}{4}D (D-2) (D-3) \rho^{ }_{\It{h}{0}}
\nn&-&\frac{1}{2} D(D-3) (2 D-7) \rho^{ }_{\It{h}{1}}+\frac{1}{4} D (D-3) (2 D+21) \rho^{ }_{\It{h}{2}}+2(2 D^2-7 D+4) \rho^{ }_{\It{h}{4'}}
\nn&+& \left(10 D^2-47 D+52\right)\rho^{ }_{\It{h}{5'}}-4 \left(D^2-2 D-2\right)\rho^{ }_{\It{h}{6'}}-2(2 D^2-7 D+4 )\rho^{ }_{\It{h}{7'}}
\nn&-& D \left(2 D^2-11 D+15\right) \rho^{ }_{\It{i}{1}} \,, \la{rhoAB}
\\ %%%%%%%%%%%%%%%
\rho^{ }_\text{(vi)} (\omega) &\equiv& -\rho^{ }_\text{(vii)} (\omega) + D (D-2) (D-3) \rho^{ }_{\It{b}{0}}+D (D-3) \rho^{ }_{\It{b}{1}}+3D(D-3) \rho^{ }_{\It{b}{2}}
\nn&+&\frac{7}{12}D(D-2)(D-3) \rho^{ }_{\It{f}{0}}+\frac{1}{4} D(D-3) \rho^{ }_{\It{f}{1}}-\frac{5}{4} D(D-2) (D-3) \rho^{ }_{\It{h}{0}}
\nn&-&\frac{3}{2} D (D-3) \rho^{ }_{\It{h}{1}}-\frac{25}{4} D (D-3) \rho^{ }_{\It{h}{2}}-2 D(D-3) \rho^{ }_{\It{h}{4'}}
\nn&-&(D-3) (7 D-12) \rho^{ }_{\It{h}{5'}}+4 D(D-3) \rho^{ }_{\It{h}{6'}}+2 D(D-3) \rho^{ }_{\It{h}{7'}}-D(D-3) \rho^{ }_{\It{i}{1}}
\nn&+&\frac{1}{4}D (D-2) (D-3) \rho^{ }_{\It{j}{0}}+2 D (D-3) \rho^{ }_{\It{j}{1}}+D(D-3) \rho^{ }_{\It{j}{2}}
\nn&+&\left(3 D^2-16 D+12\right) \rho^{ }_{\It{j}{3}}+ \frac{1}{2} \left(3 D^2-16 D+12 \right)\rho^{ }_{\It{j}{4}}
\nn&-&D(D-6) \rho^{ }_{\It{j}{5}}-2 D (D-6) \rho^{ }_{\It{j}{6} } \,, \la{rhoDH}
\\ %%%%%%%%%%%%%%%
\rho^{ }_\text{(v)} (\omega) &\equiv& -\frac{1}{2} D(D-2)^2 (D-3)\rho^{ }_{\It{b}{0}}-D(D-2) (D-3) \rho^{ }_{\It{b}{1}}-\frac{3}{2} D(D-2) (D-3)\rho^{ }_{\It{b}{2}}
\nn&+&\frac{1}{4}D (D-2)^2(D-3) \rho^{ }_{\It{d}{0}}+D(D-2) (D-3) \rho^{ }_{\It{d}{1}}+3 (D-2) (D-3) \rho^{ }_{\It{d}{2}}
\nn&+&2(D-2)^2 (D-3) \rho^{ }_{\It{d}{3}}+\frac{5}{6} D(D-2) (D-3) \rho^{ }_{\It{f}{0}}+\frac{1}{4} D(D-3)(D+6) \rho^{ }_{\It{f}{1}}
\nn&-&\frac{1}{2}D (D-2) (D-3) \rho^{ }_{\It{h}{0}}-\frac{1}{2} D(D-3)(D+6)\rho^{ }_{\It{h}{2}}+\frac{1}{4}D (D-2) (D-3)\rho^{ }_{\It{h}{3}}
\nn&-&2 (D-2)^2\rho^{ }_{\It{h}{4'}}+D(D-2) \rho^{ }_{\It{h}{4}}-(D-2) (3 D-8) \rho^{ }_{\It{h}{5'}}
\nn&+&\frac{1}{2} D(D-2) \rho^{ }_{\It{h}{5}}+4(D-2) \rho^{ }_{\It{h}{6'}}-2D(D-2 ) \rho^{ }_{\It{h}{6}}+2 (D-2)^2 \rho^{ }_{\It{h}{7'}}
\nn&-&D( D-2) \rho^{ }_{\It{h}{7}}+D(D-2) (D-3)\rho^{ }_{\It{i}{1}}-D(D-2) (D-3)\rho^{ }_{\It{i}{2}}
\nn&-&\frac{1}{2} D (D-2)(D-3) \rho^{ }_{\It{i}{3}} \,. \la{rhoEFG}
\ea
We notice that the masters $\It{h}{4'}-\It{h}{7'}$ cancel in the sum of all diagrams, which was to be expected based on their absence in the results of \cite{Zhu:2012be}. 

The reason we have chosen to write the spectral function in a form slightly different from the one used in \cite{Zhu:2012be} is that the above formulation allows us to separate all the IR sensitive masters with squared propagators to $\rho^{ }_\text{(v)} (\omega)$. This part can furthermore be written in an alternative form using the one-loop gluon polarization tensor
\ba
 \Pi_{\mu\nu}^{ab}(Q)&=& g^2\delta^{ab}N_c\bigg[ (D-2) \Tint{R}\fr{1}{R^2}\delta_{\mu\nu} 
 + 2(Q_\mu Q_\nu - Q^2\delta_{\mu\nu})\Tint{R}\fr{1}{R^2(Q-R)^2}
 \nn&-&\fr{D-2}{2}\Tint{R}\fr{(2R-Q)_\mu(2R-Q)_\nu}{R^2(Q-R)^2}\bigg] 
 \,, \la{Pi}
\ea
the transverse and longitudinal components of which read
\ba
\Pi_T^{ }(Q)\delta^{ab} = 
   \fr{1}{D-2}\left(\Pi_{\mu\mu}^{ab}(Q)-\fr{Q^2}{q^2}\Pi_{00}^{ab}(Q)\right)
  \,, &\quad&
  \Pi_E^{ }(Q)\delta^{ab} =  
  \fr{Q^2}{q^2}\Pi_{00}^{ab}(Q)
  \,. \la{PiT0} 
\ea
A straightforward exercise namely shows that one can write
\ba
\rho^{ }_\text{(v)} (\omega)  &\equiv& -\frac{D(D-2)(D-3) }{ D^2-1} \Tint{Q} \bigg\{(D-3)(D+1)\frac{\Pi _T(Q)q^2p_n{}^2}{Q^4(Q-P)^2}-2 (D-2)\frac{\Pi _T(Q)q^4}{Q^4(Q-P)^2}\bigg\}
\nn&-&\frac{D(D-2)(D-3) }{ D^2-1} \Tint{Q} \bigg\{2\frac{\Pi _T(Q)q^2}{Q^2(Q-P)^2}
+(D-1) \frac{\Pi_E(Q)q^2}{Q^2(Q-P)^2}\bigg\}
 \,, \label{newv}
\ea
where the different integrals take the forms
\ba
&&\hspace{-1.5cm} \im \biggl[
 \Tint{Q} \frac{\Pi_T(Q)q^2p_n^2}{Q^2(Q-P)^2}
 \biggr]_{P \to (-i [\omega + i 0^+],\vec{0})}
 \nn
 & = &\rho^{ }_{\It{b}{0}}+(D-1)\rho^{ }_{\It{d}{2}}-2\frac{D-1}{D-2}\rho^{ }_{\It{h}{1}}-\frac{D-1}{D-2}\rho^{ }_{\It{h}{2}}-\fr{1}{2 }\rho^{ }_{\It{f}{0}}\,,
\\
&&\hspace{-1.5cm} \im \biggl[
 \Tint{Q} \frac{\Pi_T(Q)q^4}{Q^2(Q-P)^2}
 \biggr]_{P \to (-i [\omega + i 0^+],\vec{0})}
 \nn&=&
 \fr{D-1}{D-2} \rho^{ }_{\It{b}{2}}
+\fr{D^2-1}{D}\rho^{ }_{\It{d}{3}}-\fr{D-1}{2(D-2)} \rho^{ }_{\It{f}{2}}-\fr{D^2-1}{D(D-2)}\rho^{ }_{\It{h}{5'}}-2\rho^{ }_{\It{h}{6*}} \,,
\\
&&\hspace{-1.5cm} \im \biggl[
 \Tint{Q} \frac{\Pi_T(Q)q^2}{Q^2(Q-P)^2}
 \biggr]_{P \to (-i [\omega + i 0^+],\vec{0})}
=
 (D-1)\rho^{ }_{\It{b}{2}}-\fr{1}{6}\rho^{ }_{\It{f}{0}} -3\fr{D-1}{D-2} \rho^{ }_{\It{f}{1}} \,,
\\
&&\hspace{-1.5cm} \im \biggl[
 \Tint{Q} \frac{\Pi_E(Q)q^2}{Q^2(Q-P)^2}
 \biggr]_{P \to (-i [\omega + i 0^+],\vec{0})}
=
 \fr{D-2}{6}\rho^{ }_{\It{f}{0}} +\fr{3D^2-13D+10}{2(D-2)} \rho^{ }_{\It{f}{1}} \,.
\ea 
The virtue of this formulation is the absence of masters with squared propagators, which altogether avoids the need to introduce an auxiliary mass parameter to be differentiated upon, as done in \cite{Zhu:2012be}.

The evaluation of the new primed masters utilizes the methods developed in \cite{Laine:2011xm,Zhu:2012be}, and will be explained in some detail in appendix \ref{newcases}. The independent checks we have performed for the other masters, already appearing in \cite{Zhu:2012be}, will on the other hand not be discussed further here, as they all produced positive results.

%%%%%%%%%%%%%%%%%%%%%%%%%%%%%%%%%%%%%%%%%%%%%%%%%%%%%%%%
\section{HTL resummation}\label{HTL}
%%%%%%%%%%%%%%%%%%%%%%%%%%%%%%%%%%%%%%%%%%%%%%%%%%%%%%%%

Our goal being to extend the range of applicability of the perturbative spectral function to frequencies of order $\omega\sim gT$, we will next implement in our calculation an HTL resummation that allows a consistent treatment of this parameter region. Following the logic of \cite{Laine:2011xm}, we note that one can approximate the resummed spectral function as
\be
 \rho^\rmii{QCD}_\rmii{resummed} 
 \; = \; 
 \rho^\rmii{QCD}_\rmii{resummed} 
 -  \rho^\rmii{HTL}_\rmii{resummed} 
 + \rho^\rmii{HTL}_\rmii{resummed}  
 \; \approx \;
 \rho^\rmii{QCD}_\rmii{naive} 
 -  \rho^\rmii{HTL}_\rmii{naive} 
 + \rho^\rmii{HTL}_\rmii{resummed}  
 \,, \la{master_resum}
\ee
where `QCD' refers to the full theory, while `HTL' signifies a calculation performed using the HTL effective action. In the second step above we have used the fact that the difference between the QCD and HTL spectral functions should be free from IR problems, allowing us to perform the two calculations in a `naive' form, i.e.~using expansions in powers of the coupling $g$. Of the three terms on the right hand side of eq.~(\ref{master_resum}), we have already computed the first one, so in the following we will only consider the naive and resummed HTL calculations to the necessary leading order in perturbation theory. For more details of the general procedure, we refer the interested reader to ref.~\cite{Laine:2011xm}.

The HTL calculation employs resummed gluon propagator and vertex functions, of which the former reads
\be
 \bigl\langle
  A^a_\mu(X) \, A^b_\nu(Y)
 \bigr\rangle_\text{HTL} 
 = 
 \delta^{ab}_{ }\, \Tint{Q}
 e^{i Q\cdot (X - Y)}
 \biggl[
   \frac{\mathbbm{P}^T_{\mu\nu}(Q)}{Q^2 + \Pi^\text{HTL}_T (Q)} 
  + \frac{\mathbbm{P}^E_{\mu\nu}(Q)}{Q^2 + \Pi^\text{HTL}_E (Q)}
  + \frac{\xi\, Q_\mu Q_\nu}{Q^4} 
 \biggr] 
 \,, \la{prop_HTL}
\ee
where $\xi$ denotes the gauge parameter and $\Pi^\text{HTL}$ the transverse and longitudinal HTL self energies,
\ba
   \Pi^\text{HTL}_T(Q) &=& 
  \frac{\mE^2}{D - 2}
  \int_{z} 
  \biggl( 1 - \frac{Q^2}{q^2} 
  \frac{q z}{iq_n + q z}
  \biggr) 
  \,, \quad
  \Pi^\text{HTL}_E(Q) =  
  \frac{\mE^2 Q^2}{q^2}
  \int_{z} 
  \frac{q z}{iq_n + q z}
  \,, \la{PiT} \\
 \mE^2& \equiv& g^2 \Nc (D-2)^2 \int_{r} \frac{n_r}{r} =\fr{\Nc}{3}g^2T^2
 \,, \quad \; n_q \equiv \frac{1}{{\text e}^{\beta q}-1}\,. \la{mmE}
\ea
As discussed in section 5.2 and appendix C of \cite{Laine:2011xm}, in the bulk channel the vertex function is not needed for a LO HTL calculation. In the shear case, the situation appears to be similar, as we have explicitly checked that the HTL vertex functions contribute to $\rho^\rmii{HTL}_\rmii{naive}$ at least at one order higher in $\omega/T$ than the propagator corrections. We will thus ignore the vertex part of the calculation in what follows,\footnote{This implies throwing out terms of order $\mE^2\omega^2$ from the result, which were kept in the bulk channel work of \cite{Laine:2011xm} and should in principle be accounted for in a consistent leading order HTL calculation. As including vertex corrections in the resummed HTL spectral function is, however, a technically rather tedious exercise and our present results indicate that even the more dominant HTL propagator corrections are negligible at all interesting values of $\omega$, we have decided to refrain from performing this calculation. This issue will be briefly returned to in section \ref{results}.} which leads to a rather simple result for the resummed Euclidean correlator,
\ba
% && \hspace*{-1cm} 
 \frac{\widetilde G_\eta^\rmii{HTL}(P)}{4 d_A c_\eta^2 } & = & 
 \frac{D(D-2)(D-3) }{ 2\left(D^2-1\right)}\Tint{Q}\biggl\{(D+1)(D-3)\frac{q^2p_n{}^2}{\Delta_T(Q)\Delta_T(Q-P)}
 \nn &-&  2(D-2)\frac{q^4}{\Delta_T(Q)\Delta_T(Q-P)}+4 \frac{q^2Q^2}{\Delta _T(Q)\Delta_T(Q-P)}
 \nn&-&\left(D^2-2D-1\right)\frac{Q^2(Q-P)^2}{\Delta_T(Q)\Delta_T(Q-P)}-2\frac{Q^2(Q-P)^2}{\Delta_E(Q)\Delta_E(Q-P)}\biggr\}
 \nn &+&\frac{D(D-2)(D-3) }{ (D+1)}\Tint{Q} \biggl\{\frac{q^2(Q-P)^2}{\Delta_T(Q)\Delta _E(Q-P)}-\frac{Q^2(Q-P)^2}{\Delta_T(Q)\Delta_E(Q-P)}\biggr\}
 \,, \la{Geta_HTL} 
\ea 
with 
$
 \Delta^{ }_T(Q) \equiv Q^2 + \Pi^\text{HTL}_T(Q)
$
and
$
 \Delta^{ }_E(Q) \equiv Q^2 + \Pi^\text{HTL}_E(Q)
$.

%%%%%%%%%%%%%%%%%%%%%%%%%%%%%%%%%%%%%%%%%%%%%%%%%%%%%
\subsection{Naive calculation}\label{htl_naive}
%%%%%%%%%%%%%%%%%%%%%%%%%%%%%%%%%%%%%%%%%%%%%%%%%%%%%
Looking first into the naive HTL calculation, we proceed to expand the expression in eq.~(\ref{Geta_HTL}) in powers of the coupling, leading us to
\ba
% && \hspace*{-1cm} 
 \frac{\widetilde G_\eta^\rmii{HTL}(P)}{4 d_A c_\eta^2 } & = & 
-\frac{D(D-2)(D-3) }{ D^2-1} \Tint{Q} \bigg\{(D-3)(D+1)\frac{\Pi^\text{HTL} _T(Q)q^2p_n{}^2}{Q^4(Q-P)^2}
\nn&-&2 (D-2)\frac{\Pi^\text{HTL} _T(Q)q^4}{Q^4(Q-P)^2}\bigg\}
+\frac{D(D-2)(D-3) }{ D^2-1} \Tint{Q} \bigg\{D(D-3)\frac{\Pi^\text{HTL} _T(Q)q^2}{Q^2(Q-P)^2}
\nn&-& (D-1) \frac{\mE^2q^2}{Q^2(Q-P)^2}
-(D-2)\frac{q^4}{Q^2(Q-P)^2}\bigg\}
\nn&+&\frac{D(D-2)(D-3)}{2 (D-1)} \Tint{Q} \left\{(D-3)\frac{q^2p_n{}^2}{Q^2(Q-P)^2}\right\}
 \,. \la{Geta_naive} 
\ea
The evaluation of the sum-integrals appearing here is a rather straightforward exercise that utilizes the machinery developed in \cite{Laine:2011xm} and is explained in some detail in appendix \ref{htl1}. The result of this procedure reads
\be
\frac{\rho^{\text{HTL}}_\eta(\omega)}{4d_A}\bigg|_\text{naive}=-\frac{1}{4\pi}\bigl( 1 + 2 n_{\frac{\omega}{2}} \bigr)\Bigg\{\frac{\omega^4}{10}+\frac{\mE^2}{45} \bigg[\pi^2 \omega T+\omega^2\Big(12+\frac{3}{2}\ln2-\fr{15}{2}\ln\fr{\omega}{T}\Big)\bigg]\Bigg\}\; . \label{result1x}
\ee

%%%%%%%%%%%%%%%%%%%%%%%%%%%%%%%%%%%%%%%%%%%%%%%%%%%%%
\subsection{Resummed calculation}\label{htl_full}
%%%%%%%%%%%%%%%%%%%%%%%%%%%%%%%%%%%%%%%%%%%%%%%%%%%%%
To prepare for the resummed version of the HTL calculation, we first write eq.~\nr{Geta_HTL} in the alternative form (cf.~the discussion in section 5.4 of \cite{Laine:2011xm}) 
\ba
% && \hspace*{-1cm} 
 \frac{\widetilde G_\eta^\rmii{HTL}(P)}{4 d_A c_\eta^2 } & = & 
 \frac{D(D-2)(D-3) }{ 2\left(D^2-1\right)}\Tint{Q}\biggl\{(D+1)(D-3)\frac{q^2p_n{}^2}{\Delta_T(Q)\Delta_T(Q-P)}
 \nn &-&  2(D-2)\frac{q^4}{\Delta_T(Q)\Delta_T(Q-P)}-4 \frac{q^2\Pi^\text{HTL}_T(Q)}{\Delta _T(Q)\Delta_T(Q-P)}
  \la{Geta_HTL1} \\
&-&\left(D^2-2D-1\right)\frac{\Pi^\text{HTL}_T(Q)\Pi^\text{HTL}_T(Q-P)}{\Delta_T(Q)\Delta_T(Q-P)}-2\frac{\Pi^\text{HTL}_E(Q)\Pi^\text{HTL}_E(Q-P)}{\Delta_E(Q)\Delta_E(Q-P)}\biggr\}
 \nn &-&\frac{D(D-2)(D-3) }{ (D+1)}\Tint{Q} \biggl\{\frac{q^2\Pi^\text{HTL}_E(Q-P)}{\Delta_T(Q)\Delta _E(Q-P)}+\frac{\Pi^\text{HTL}_T(Q)\Pi^\text{HTL}_E(Q-P)}{\Delta_T(Q)\Delta_E(Q-P)}\biggr\}\,, \nonumber 
\ea
where we have dropped a number of uninteresting contact terms, polynomial in the external momentum. Making then use of relations of the type
\ba
 & & \hspace*{-2cm}
 \im \biggl\{ 
   T \sum_{q_n} \frac{1}{\Delta^{ }_T(Q)\Delta^{ }_T(Q-P)}
 \biggr\}_{P \to (-i [\omega + i 0^+],\vec{0})}
 \nn 
 & = &  
 \bigl( 1 + 2 n_{\frac{\omega}{2}} \bigr)
 \int_{-\infty}^{\infty} \! \frac{{\rm d}q^0}{\pi} \, 
 \, \rho^{ }_T(q^0,q) \, \rho^{ }_T(\omega - q^0,q) 
 \, \frac{n_{q^0} n_{\omega - q^0} }{n^2_{\frac{\omega}{2}} }
 \,, \la{rho_rho}
\ea
where we have set $D=4$ and denoted $\rho_T(q_0,q) \equiv {\rm Im} \{1/\Delta_T(q_n,q)\}_{q_n \to -i[q_0+i0^+]}$ (see appendix \ref{htl2} for an exhaustive list of definitions and relations of this kind), the spectral function obtains the compact form ($(\mathcal{P-Q})^2\equiv (\omega-q_0)^2-q^2$)
\ba
 && \hspace*{-2cm} 
 \frac{\rho_\eta^\rmii{HTL}(\omega)}{4 d_A c_\eta^2 }  =  
 -\frac{32 \bigl( 1 + 2 n_{\frac{\omega}{2} } \bigr)}{15(4\pi)^2}
 \int_0^\infty \! {\rm d}q \, q^2 \int_{-\infty}^{\infty} \! \fr{{\rm d}q^0}{\pi} \, \Bigg\{\nn
 &\times&\bigg[ \Bigl(5\omega^2q^2+4q^4+\mathcal{Q}^2(4q^2+7(\mathcal{P-Q})^2 ) \Bigr) \rho^{ }_T(q^0,q) \rho^{ }_T(\omega - q^0,q)
 \nn &&  
 +2 \mathcal{Q}^2(\mathcal{P-Q})^2 \rho^{ }_E(q^0,q) \rho^{ }_E(\omega - q^0,q) 
 \nn 
 && + 6\Big(q^2(\mathcal{P-Q})^2 + \mathcal{Q}^2(\mathcal{P-Q})^2\Bigr) \rho^{ }_T(q^0,q) \rho^{ }_E(\omega - q^0,q) \bigg]
 \nn &\times& \frac{n^{ }_{q^0}n^{ }_{\omega - q^0}}{n^2_{\frac{\omega}{2}}}
 \Bigg\}
 \,. \la{rho_HTL_full} 
\ea
All terms appearing here are of the general types already encountered in \cite{Laine:2011xm}, and we thus refer the interested reader to appendix C.3 of this reference for details of their evaluation. The only difference between that calculation and ours is that due to the more UV divergent form of our integrals, we are not able to replace the distribution functions by their classical limits, ${n^{ }_{q^0}n^{ }_{\omega - q^0}} / {n^2_{\frac{\omega}{2}}} \to \frac{\omega^2}{4 q^0(\omega - q^0)}$, in the numerical evaluation of the integrals. This implies that our result will explicitly depend on three scales, $\omega$, $T$ and $\mE$.

Following ref.~\cite{Laine:2011xm}, we finally write the result of the resummed HTL calculation as
\be
\frac{\rho^{\text{HTL}}_\eta(\omega)}{4d_A}\bigg|_\text{resummed}=\frac{\rho^{\text{HTL}}_\eta(\omega)}{4d_A}\bigg|_\text{naive}+\frac{\mE^4}{4\pi}\bigl( 1 + 2 n_{\frac{\omega}{2}} \bigr)\phi_{\text{HTL}}^\eta(\omega/T,\mE/T)\, , \label{resresult1}
\ee
where we have for convenience separated out the naive HTL result from the rest.

%%%%%%%%%%%%%%%%%%%%%%%%%%%%%%%%%%%%
\section{Results}\label{results}
%%%%%%%%%%%%%%%%%%%%%%%%%%%%%%%%%%%%%%%

%%%%%%%%%%%%%%%%%%%%%%%%%%%%%%%%%%%%%%%%%%%%
\begin{figure}
\centering
\epsfxsize=7cm \epsfbox{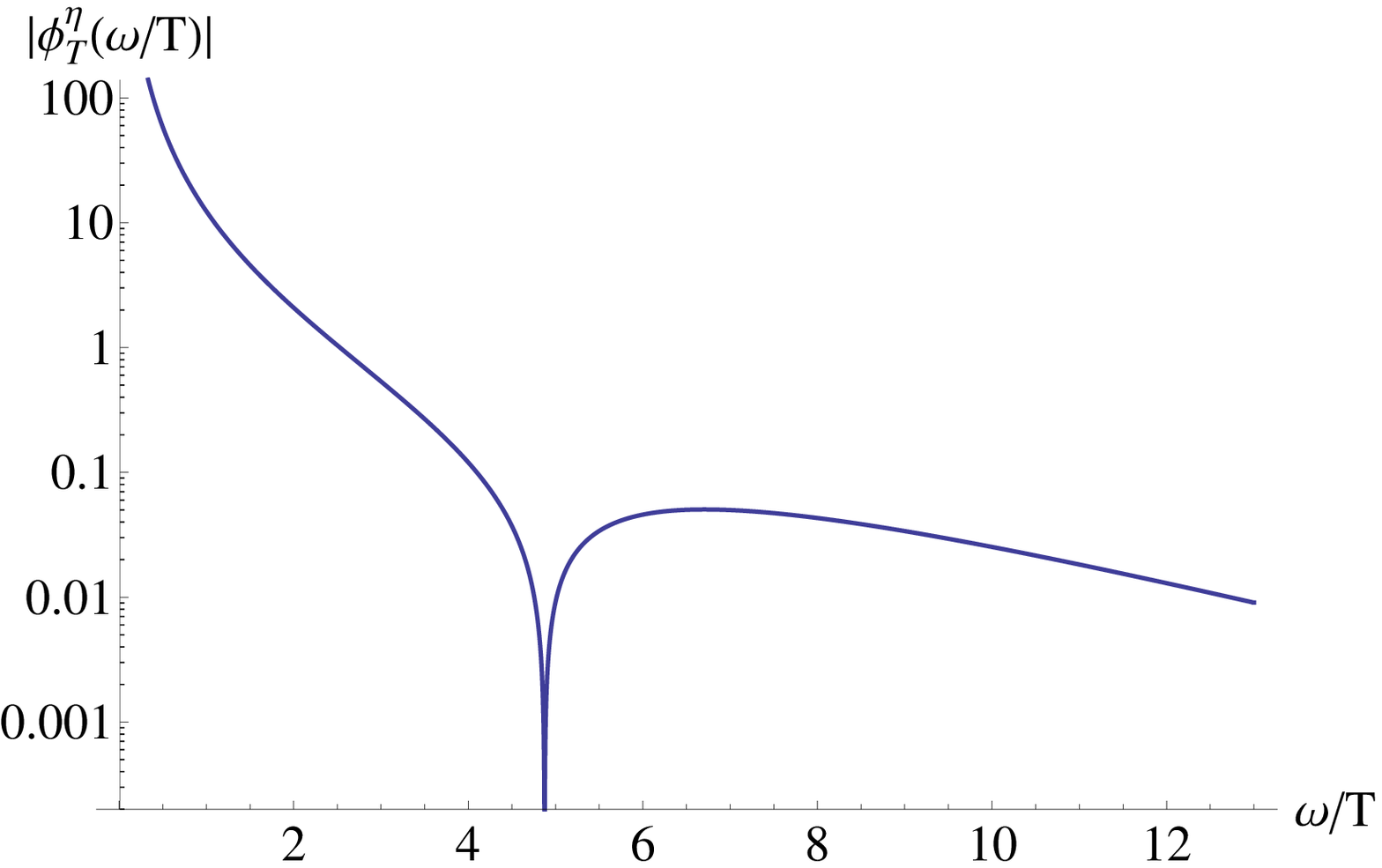}$\;\;\;$\epsfxsize=7.8cm \epsfbox{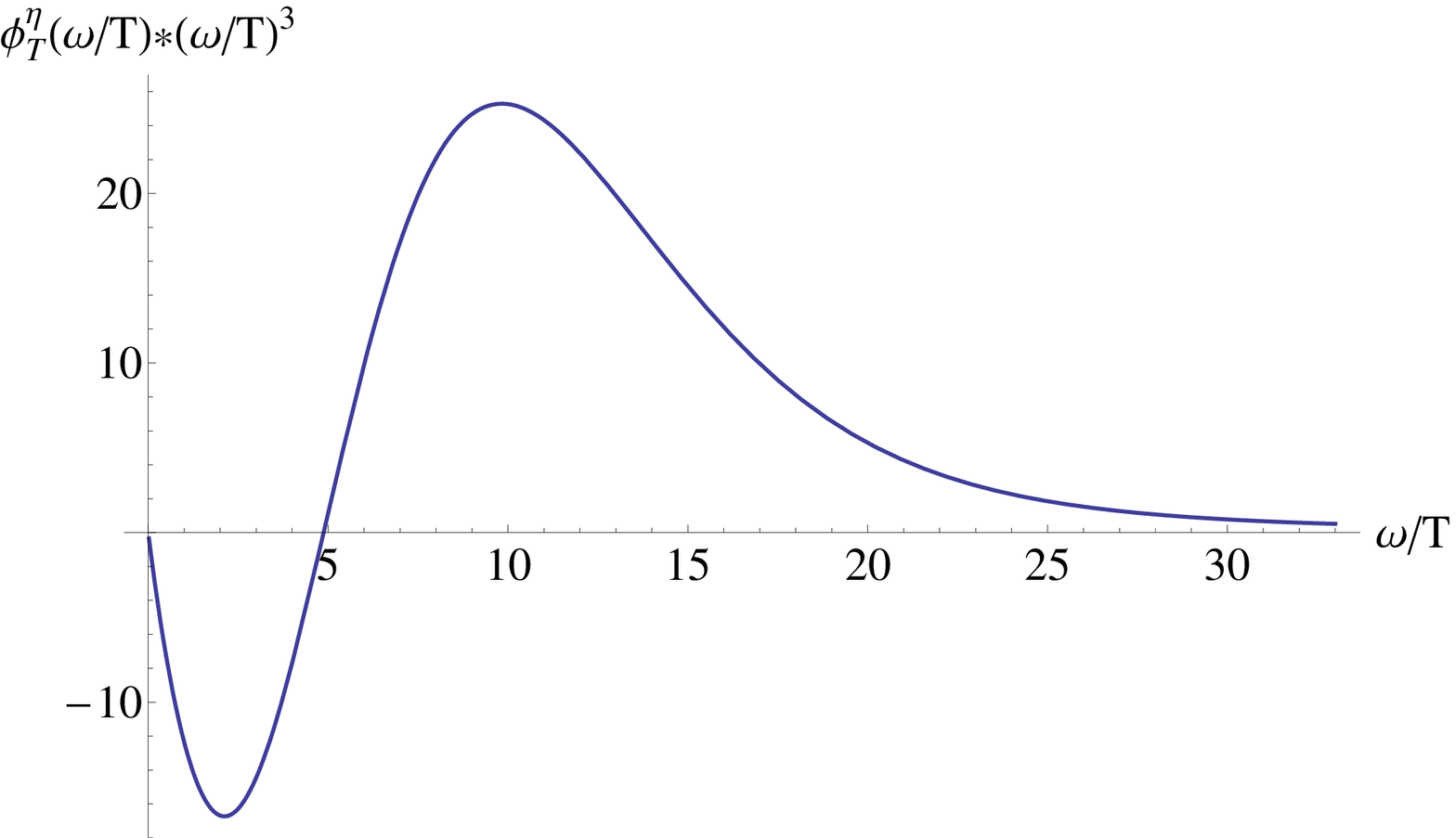}
\caption{The behavior of the function $\phi_T^\eta(\omega/T)$ on a linear and a logarithmic scale, multiplied by $(\omega/T)^3$ in the latter case.}
\label{res1}
\end{figure}
%%%%%%%%%%%%%%%%%%%%%%%%%%%%%%%%%%%%%%%%%%%%

Having now discussed the different ingredients involved in our calculation, we will next collect, display and analyze our results. For clarity, this section is divided into three parts: First, we collect the unresummed result, highlighting the fact that in the course of our work we discovered and corrected a subtle technical error in our original work on the same subject \cite{Zhu:2012be}. Then, we employ the HTL resummation in the form described above and study how this affects the spectral function, while in the final third part of the section we apply our results to the evaluation of the imaginary time correlator and also briefly discuss the shear sum rule.

\subsection{Unresummed spectral function \label{unres}}

As demonstrated already in \cite{Zhu:2012be}, the unresummed NLO shear spectral function can be written in the form
\ba
\frac{\rho_\eta(\omega)}{4d_A}&=&\frac{\omega^4}{4\pi}\bigl( 1 + 2 n_{\frac{\omega}{2}} \bigr)\Bigg\{-\frac{1}{10}+\frac{g^2N_c}{(4\pi)^2}\bigg(\frac{2}{9}+\phi_T^\eta(\omega/T)\bigg)\Bigg\}\; , \label{result1}
\ea
where we have defined a dimensionless function $\phi_T^\eta(\omega/T)$ using the quantities $\rho^{ }_\text{(n)} (\omega)$ introduced in section \ref{unresummed}. This function was first determined already in \cite{Zhu:2012be}, but to our surprise, our new formulation via eq.~(\ref{newv}) was seen to lead to a different result, exhibiting a less divergent behavior at small $\omega$. This discrepancy was settled after a lengthy struggle upon the discovery of a subtle computational error in those masters of \cite{Zhu:2012be} that contain a squared propagator. This issue, which boils down to a number of UV-suppressed analytic contributions having been missed in \cite{Zhu:2012be}, is explained thoroughly in appendix \ref{discre} below.

Having settled the discrepancy, we display the behavior of the corrected $\phi_T^\eta(\omega/T)$ function in fig.~\ref{res1}. A comparison with the corresponding fig.~2 of ref.~\cite{Zhu:2012be} reveals that the difference between the two results vanishes quickly at large $\omega$, yet becomes qualitatively important in the IR region. In particular, we see that while the analytic large-$\omega$ result
\ba
\phi_T^\eta(\omega/T)&=&\frac{41\pi^6T^6}{3\omega^6} +{\mathcal O}(T^8/\omega^8)\; ,
\ea
still holds --- in accordance with the arguments of \cite{CaronHuot:2009ns} --- the correct $\omega \to 0$ limit of the quantity is now of order $T^2/\omega^2$ (the constant of proportionality here is not known analytically, nor does it have any specific physical significance).  This has the important consequence that the spectral function itself has now a linear behavior at low $\omega$, implying that it can be straightforwardly used in the evaluation of the integrals appearing in the imaginary time correlator and the shear sum rule (cf.~the discussion of this issue in \cite{Zhu:2012be}).

%%%%%%%%%%%%%%%%%%%%%%%%%%%%%%%%%%%%%%%%%%%%
\begin{figure}
\centering
\epsfxsize=8.5cm \epsfbox{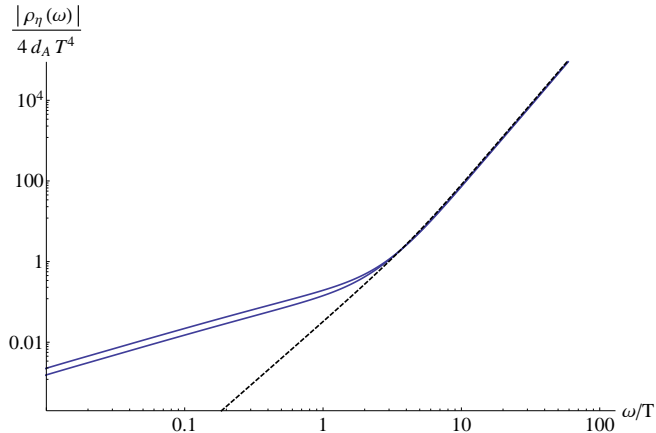}
\caption{The behavior of the absolute value of the (negative) shear spectral function for $T=3T_c$ (corresponding to $3.75\Lambda_\tinymsbar$). The two blue curves stand for the NLO result evaluated with $\bar{\Lambda}=0.5\bar{\Lambda}_\text{opt}$ and $\bar{\Lambda}=2\bar{\Lambda}_\text{opt}$, while the dashed black curve shows the leading order (LO) result.}
\label{res2}
\end{figure}
%%%%%%%%%%%%%%%%%%%%%%%%%%%%%%%%%%%%%%%%%%%%

Next, we insert the numerical function $\phi_T^\eta(\omega/T)$ into eq.~(\ref{result1}) in order to study the behavior of the spectral function numerically. Using the familiar one-loop result for $\alpha_s$, setting $N_c=3$ and choosing the renormalization scale $\bar{\Lambda}$ to be varied by a factor of 2 around the `EQCD value' \cite{Kajantie:1997tt}\footnote{As discussed in \cite{Zhu:2012be}, at large energies it might seem more natural to choose $\bar{\Lambda}$ to be proportional to $\omega$. This, however, would only have a minuscule impact on the results.}
\ba
\ln\,\frac{\bar{\Lambda}_\text{opt}}{4\pi T}&=&-\gamma_\text{E}-\frac{1}{22} \; ,
\ea
we obtain the behavior shown in fig.~\ref{res2}. Comparing this to the results of \cite{Zhu:2012be}, we again observe a fast approach of the result towards the free theory limit in the UV, but an important difference is that this time the spectral function does not change sign at small $\omega$, but stays negative at all frequencies.

\subsection{HTL resummed spectral function \label{HTLresults}}

Even though we could now directly proceed to use the corrected unresummed spectral function in the applications described above, it is interesting to also study the impact of the HTL resummation on its IR behavior, as this is expected to expand the region of validity of the result to frequencies of order $gT$. Here, our treatment is both strongly motivated by and closely follows the bulk computation of ref.~\cite{Laine:2011xm}, where it was seen that performing a one-loop HTL resummation was enough to turn the leading ${\mathcal O}(\omega)$ IR behavior of the spectral function to an ${\mathcal O}(\omega^2)$ one.

%%%%%%%%%%%%%%%%%%%%%%%%%%%%%%%%%%%%%%%%%%%%
\begin{figure}
\centering
\epsfxsize=7.8cm \epsfbox{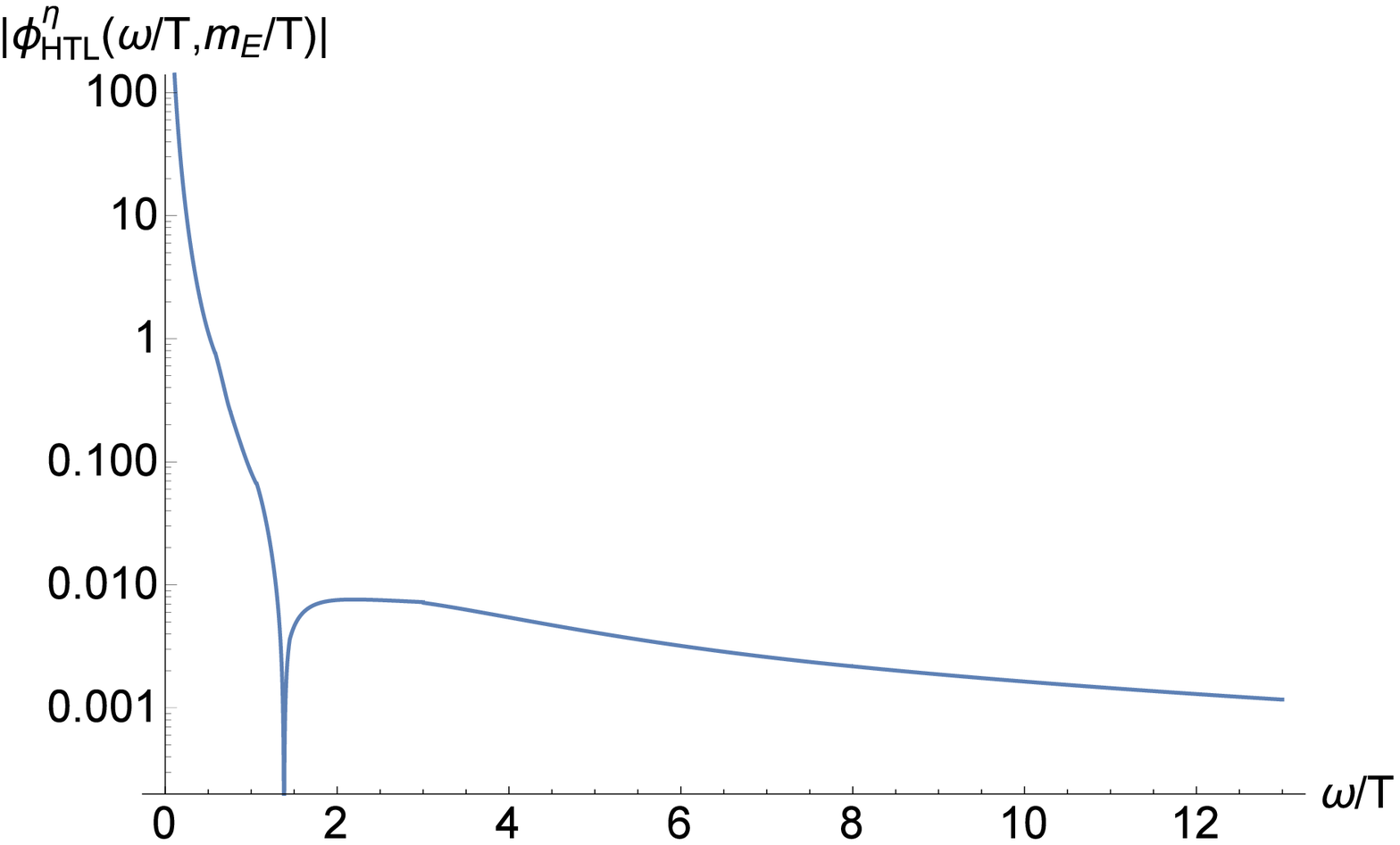}$\;\;\;$\epsfxsize=7cm \epsfbox{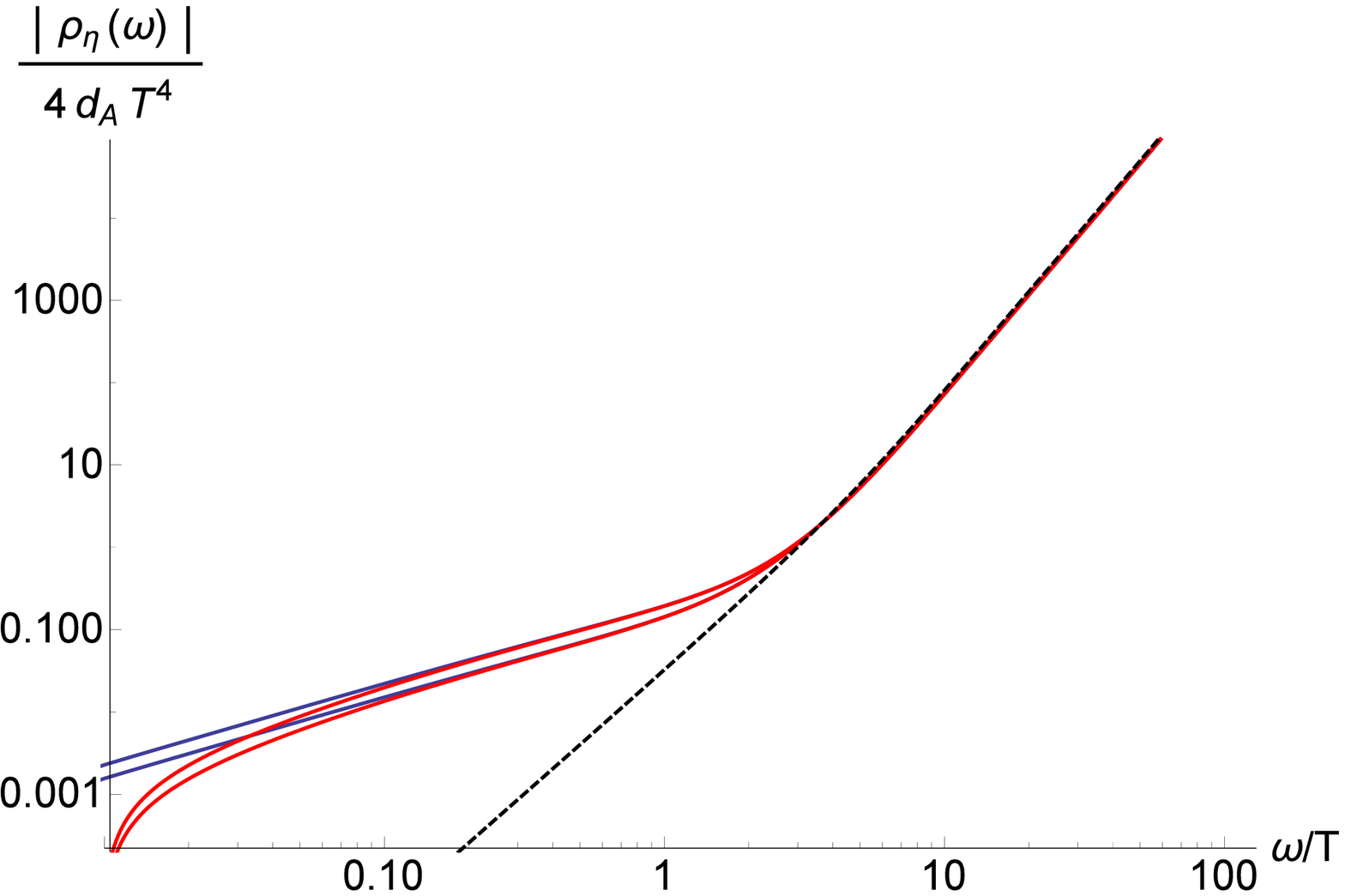}
\caption{Left: The behavior of the function $\phi_\text{HTL}^\eta(\omega/T,\mE/T)$ on a logarithmic scale, with its value turning from positive to negative at around $\omega\approx 1.38T$. Right: The absolute value of the (negative) resummed shear spectral function $\rho_\eta^\text{res}(\omega)/(4d_A)$ (red curves), compared with the unresummed NLO result already displayed in fig.~\ref{res2} (blue curves) and the LO result (dashed black curve). For the NLO results, the two curves again correspond to the renormalization scale choices $\bar{\Lambda}=0.5\bar{\Lambda}_\text{opt}$ and $\bar{\Lambda}=2\bar{\Lambda}_\text{opt}$.}
\label{fig:phiHTL}
\end{figure}
%%%%%%%%%%%%%%%%%%%%%%%%%%%%%%%%%%%%%%%%%%%%

Collecting the results of the previous two sections, we see that the resummed shear spectral function obtains the form
\ba
\frac{\rho_\eta^\text{res}(\omega)}{4d_A}&=&\frac{\omega^4}{4\pi}\bigl( 1 + 2 n_{\frac{\omega}{2}} \bigr)\Bigg\{-\frac{1}{10}+\frac{g^2N_c}{(4\pi)^2}\bigg(\frac{2}{9}+\phi_T^\eta(\omega/T)\bigg)+\frac{\mE^4}{\omega^4}\,\phi_{\text{HTL}}^\eta(\omega/T,\mE/T)\Bigg\}\, , \nonumber \\ \label{resultHTL1}
\ea
where the term linear in $\omega$ in the naive HTL result (\ref{result1x}) is not visible due to the way we chose to write the resummed HTL contribution in eq.~(\ref{resresult1}). It, however, remains to be shown that the function $\phi_{\text{HTL}}^\eta(x,\mE/T)$ indeed starts with an $x^2$ term in the IR limit, so that the problems we have just solved with our new unresummed calculation have not returned. This turns out to be a somewhat nontrivial exercise, as the function now depends on the extra dimensionless parameter $\mE/T$ due to the fact that we were not able to replace the Bose-Einstein distribution functions by their classical limits in its evaluation (cf.~the discussion in the previous section). Inspecting the behavior of eq.~(\ref{rho_HTL_full}) in detail, we see that the required linear term appears in eq.~(\ref{resresult1}) in the limit where there is a clear hierarchy between the scales $T$ and $\mE$, $\mE \ll T$. There, one can show even analytically that the function $\phi_{\text{HTL}}^\eta(x,\mE/T)$ indeed behaves like $x^2$ in the small-$x$ limit.

At phenomenologically interesting temperatures slightly above the critical temperature of the deconfinement transition, $T_c$, there is unfortunately no hierarchy between the scales $T$ and $\mE$, and thus no appearance of a term linear in $\omega$ in eq.~(\ref{rho_HTL_full}). The effects of this are visible in fig.~\ref{fig:phiHTL}, where we display the behaviors of both $\phi_{\text{HTL}}^\eta(\omega/T,\mE/T)$ and the entire $\rho_\eta^\text{res}(\omega)$ at $T=3T_c$. We observe that the HTL contribution has a visible effect on the result only at very small $\omega$, but that there it again turns the result more IR divergent, and in fact makes the spectral function approach a constant for very small values of $\omega$. 

As discussed already in section \ref{HTL}, our current resummation does not represent a complete LO HTL calculation, as it does not include contributions from the HTL vertex function. We have, however, explicitly verified that the vertex contribution to eq.~(\ref{result1x}) does not include a term proportional to $\mE^2\omega T$, but starts at earliest at order $\mE^2\omega^2$. Considering also the fact that our present HTL result shows perfect numerical agreement with the unresummed one down to frequencies $\omega \approx 0.1 T$, we find it not worth the effort to undertake the challenge of evaluating the resummed HTL vertex contribution to the shear spectral function.\footnote{Note, however, that for frequencies $\omega\sim gT$, the neglected terms are still formally of ${\mathcal O}(1)$ in comparison with terms included in $\phi_T^\eta(\omega/T)$.} In fact, due to the more bening small-$\omega$ behavior of the unresummed result, we will choose to use it in the evaluation of the imaginary time correlator below.

\subsection{Sum rule and imaginary time correlator}

%%%%%%%%%%%%%%%%%%%%%%%%%%%%%%%%%%%%%%%%%%%%
\begin{figure}
\centering
\epsfxsize=8.5cm \epsfbox{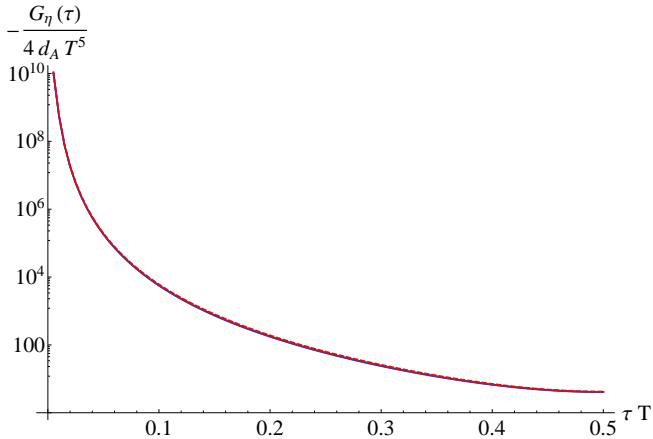}
\caption{The imaginary time correlator $G_\eta^\text{def}(\tau)$ of eq.~(\ref{tau1}) displayed on a logarithmic scale. Just as in fig.~\ref{res2}, altogether three functions are displayed here: Two blue NLO curves and one dashed red LO one.}
\label{res3}
\end{figure}
%%%%%%%%%%%%%%%%%%%%%%%%%%%%%%%%%%%%%%%%%%%%

Motived by the above considerations, we will now apply the unresummed shear spectral function of section \ref{unres} to the determination of the corresponding imaginary time correlator and to a discussion of the shear sum rule (see e.g.~\cite{Romatschke:2009ng,Meyer:2010gu,Schroder:2011ht}), 
\ba
-\fr{1}{16\pi}\int_{-\infty}^{\infty}\fr{{\rm d}\omega}{\omega} \Big\{\rho_\eta(\omega)-\rho_\eta(\omega)|_{T=0}\Big\}
&=&\fr{2}{3}e(T)+{\mathcal O}(g^4)\,. \label{specin}
\ea
The linear IR behavior of our new result makes these two tasks in principle feasible, but the latter is still plagued by the existence of undetermined contributions to the spectral function of the form $\omega\delta(\omega)$, which contribute to integrals of $\rho_\eta(\omega)/\omega$. For the sake of curiosity, we have nevertheless performed the integral on the left-hand side of eq.~(\ref{specin}) and found that even at order $g^2$ the result is within a few per cent of the energy density residing on the right-hand side. This indicates that the so far unknown contact terms in the spectral function should be numerically subleading.

With the imaginary time corerlator, the story is significantly simpler, as the delta function terms only contribute uninteresting constant terms to the integral
\begin{equation}
 G_\eta(\tau) =
 \int_{0}^\infty
 \frac{{\rm d}\omega}{\pi} \rho_\eta(\omega)
 \frac{\cosh\Big[\! \left(\frac{\beta}{2} - \tau\right)\omega\Big]}
 {\sinh\frac{\beta \omega}{2}}\, ,\quad \quad 0<\tau <\beta \, . \label{tau1}
\end{equation}
Plugging our unresummed spectral function to this expression, we obtain the behavior displayed in fig.~\ref{res3}. We observe an almost perfect agreement of the LO and NLO results, and the renormalization scale dependence of the latter is furthermore nearly invisible. This behavior is in practice identical to the results obtained in \cite{Zhu:2012be} using an IR cut-off, leaving all earlier comparisons with corresponding lattice and AdS results (see e.g.~\cite{Kajantie:2013gab,Krssak:2013jla}) unchanged.

%%%%%%%%%%%%%%%%%%%%%%%%%%%%%%%%%%%%%%%
\section{Conclusions}\label{conclusions}
%%%%%%%%%%%%%%%%%%%%%%%%%%%%%%%%%%%%%%%

The calculations reported in the paper at hand are a direct continuation of the work performed earlier in \cite{Zhu:2012be}. In this reference, we presented an NLO computation of the shear spectral function in pure SU($N$) Yang-Mills theory, which was carried out without any IR resummations. While consistent with all known limits (constraining primarily its UV behavior), this result had one surprising feature: It suggested that the perturbative spectral function approaches a constant in the small-$\omega$ limit, invalidating its direct use in the evaluation of the imaginary time shear correlator or in sum rules. This prompted us to perform a further investigation of the quantity --- in particular its IR behavior --- by performing a leading order HTL resummation, expected to extend the validity of the result to frequencies of order $gT$.

What we discovered in the course of our work was somewhat surprising: Both the naive and resummed HTL contributions to the spectral function turned out to produce terms constant in the $\omega\to 0$ limit, but in the presence of a hierarchy between the temperature and Debye mass scales, $T\gg \mE$, these exactly cancel each other on the right hand side of eq.~(\ref{master_resum}). Puzzled by this observation, we carried out a careful and fully independent check of the computation first performed in \cite{Zhu:2012be}, eventually discovering a subtle mistake there, described in detail in appendix \ref{discre}. Correcting for this was finally seen to exactly cancel the constant IR term from the unresummed spectral function, while leaving its earlier, correct UV behavior intact.\footnote{The leading UV limit of the correlator had been predicted several years earlier in \cite{CaronHuot:2009ns}.} 

Having obtained the correct unresummed and HTL resummed spectral functions, we proceeded to study their behavior as well as their effect on the imaginary time shear correlator. We observed that the HTL resummation only affects the spectral function at very small $\omega$, and that the deviation of our new imaginary time correlator from the one derived in \cite{Zhu:2012be} is in practice negligible. These results can be interpreted as reflecting the remarkably good convergence properties of the shear channel Green's functions, and make us confident that we now have the behavior of the perturbative shear spectral function under good numerical control for a wide range of frequencies. 

With the new results at hand, the main ingredient needed before we can attempt a first principles extraction of the shear viscosity of SU($N$) Yang-Mills theory is obtaining accurate continuum extrapolated lattice data for the imaginary time correlator. We hope that this challenge will be tackled by several lattice groups in the near future, followed by coordinated lattice and perturbation theory efforts in the final analytic continuation.

%%%%%%%%%%%%%%%%%%%%%%%%%%%%%%%%%%%%%%%%%%%%
\section*{Acknowledgments}

We are grateful to Mikko Laine for the many fruitful discussions we have had throughout the course of our work. The work of A.V.~was supported by the Academy of Finland, grant Nr.~273545, and that of Y.Z.~by the European Research Council grant  HotLHC, No.~ERC-2001-StG-279579.

%%%%%%%%%%%%%%%%%%%%%%%%%%%%%%%%%%%%%%%%%%%%%

%\clearpage

%%%%%%%%%%%%%%%%%%%%%%%%%%%%%%%%%
\begin{appendix}
\section{Master integrals in the unresummed calculation}\label{masters}
%%%%%%%%%%%%%%%%%%%%%%%%%%%%%%%%%%

\subsection{Definitions}

The master integrals appearing in our unresummed expression for the shear spectral function, eq.~(\ref{rhoeta1}), are defined by
\begin{align}
 \mathcal{J}_\rmi{b}^0 & \equiv
 \Tint{Q} \frac{P^4}{Q^2(Q-P)^2}
 \,, \la{Jb0} \displaybreak[0] \\
 \mathcal{J}_\rmi{b}^1 & \equiv
 \Tint{Q} \frac{P^2}{Q^2(Q-P)^2}P_T(Q)
 \,, \la{Jb1} \displaybreak[0] \\
 \mathcal{J}_\rmi{b}^2 & \equiv
 \Tint{Q} \frac{1}{Q^2(Q-P)^2}P_T(Q)^2
 \,, \la{Jb2}\displaybreak[0] \\
 \mathcal{I}_\rmi{b}^0 & \equiv
 \Tint{Q,R} \frac{P^2}{Q^2R^2(R-P)^2}
 \,,\displaybreak[0] \\
 \mathcal{I}_\rmi{b}^1 & \equiv
 \Tint{Q,R} \frac{1}{Q^2R^2(R-P)^2}P_T(Q)
 \,,\displaybreak[0] \\
 \mathcal{I}_\rmi{b}^2 & \equiv
 \Tint{Q,R} \frac{1}{Q^2R^2(R-P)^2}P_T(R)
 \,,\displaybreak[0] \\
 \mathcal{I}_\rmi{d}^0 & \equiv
 \Tint{Q,R} \frac{P^4}{Q^2R^4(R-P)^2}
 \,, \displaybreak[0] \\
 \mathcal{I}_\rmi{d}^1 & \equiv
 \Tint{Q,R} \frac{P^2}{Q^2R^4(R-P)^2}P_T(Q)
 \,,\displaybreak[0] \\
 \mathcal{I}_\rmi{d}^2 & \equiv
 \Tint{Q,R} \frac{P^2}{Q^2R^4(R-P)^2}P_T(R)
 \,,\displaybreak[0] \\
 \mathcal{I}_\rmi{d}^3 & \equiv
 \Tint{Q,R} \frac{1}{Q^2R^4(R-P)^2}P_T(R)^2
 \,,\displaybreak[0] \\
\mathcal{I}_\rmi{f}^0 & \equiv
 \Tint{Q,R} \frac{P^2}{Q^2(Q-R)^2(R-P)^2}
 \,, \displaybreak[0] \\
 \mathcal{I}_\rmi{f}^1 & \equiv
 \Tint{Q,R} \frac{1}{Q^2(Q-R)^2(R-P)^2}P_T(Q)
 \,,\displaybreak[0] \\
\mathcal{I}_\rmi{h}^0 & \equiv
 \Tint{Q,R} \frac{P^4}{Q^2R^2(Q-R)^2(R-P)^2}
 \,, \la{def_Ih0} \displaybreak[0] \\
\mathcal{I}_\rmi{h}^1 & \equiv
 \Tint{Q,R} \frac{P^2}{Q^2R^2(Q-R)^2(R-P)^2}P_T(Q)
 \,, \la{def_Ih1} \displaybreak[0] \\
\mathcal{I}_\rmi{h}^2 & \equiv
 \Tint{Q,R} \frac{P^2}{Q^2R^2(Q-R)^2(R-P)^2}P_T(R)
 \,, \la{def_Ih2} \displaybreak[0] \\
\mathcal{I}_\rmi{h}^3 & \equiv
 \Tint{Q,R} \frac{P^4}{Q^2R^4(Q-R)^2(R-P)^2}P_T(R)
 \,, \la{def_Ih3} \displaybreak[0] \\
\mathcal{I}_\rmi{h}^4 & \equiv
 \Tint{Q,R} \frac{P^2}{Q^2R^4(Q-R)^2(R-P)^2}P_T(Q)^2
 \,, \la{def_Ih4} \displaybreak[0] \\
\mathcal{I}_\rmi{h}^{4'} & \equiv
 \Tint{Q,R} \frac{1}{Q^2R^2(Q-R)^2(R-P)^2}P_T(Q)^2
 \,, \la{def_Ih4p} \displaybreak[0] \\
\mathcal{I}_\rmi{h}^5 & \equiv
 \Tint{Q,R} \frac{P^2}{Q^2R^4(Q-R)^2(R-P)^2}P_T(R)^2
 \,, \la{def_Ih5} \displaybreak[0] \\
\mathcal{I}_\rmi{h}^{5'} & \equiv
 \Tint{Q,R} \frac{1}{Q^2R^2(Q-R)^2(R-P)^2}P_T(R)^2
 \,, \la{def_Ih5p} \displaybreak[0] \\
\mathcal{I}_\rmi{h}^6 & \equiv
 \Tint{Q,R} \frac{P^2}{Q^2R^4(Q-R)^2(R-P)^2}P_T(Q)P_T(R)
 \,, \la{def_Ih6} \displaybreak[0] \\
\mathcal{I}_\rmi{h}^{6'} & \equiv
 \Tint{Q,R} \frac{1}{Q^2R^2(Q-R)^2(R-P)^2}P_T(Q)P_T(R)
 \,, \la{def_Ih6p} \displaybreak[0] \\
\mathcal{I}_\rmi{h}^{6*} & \equiv
 \Tint{Q,R} \frac{q^2r^2}{Q^2R^2(Q-R)^2(R-P)^2}
 \,, \la{def_Ih6st} \displaybreak[0] \\
\mathcal{I}_\rmi{h}^7 & \equiv
 \Tint{Q,R} \frac{P^2}{Q^2R^4(Q-R)^2(R-P)^2}P_T(Q)P_T(Q-R)
 \,, \la{def_Ih7} \displaybreak[0] \\
\mathcal{I}_\rmi{h}^{7'} & \equiv
 \Tint{Q,R} \frac{1}{Q^2R^2(Q-R)^2(R-P)^2}P_T(Q)P_T(Q-R)
 \,, \la{def_Ih7p} \displaybreak[0] \\
 \mathcal{I}_\rmi{i}^0 & \equiv
 \Tint{Q,R} \frac{(Q-P)^4}{Q^2R^2(Q-R)^2(R-P)^2}
 \,,\displaybreak[0] \\
 \mathcal{I}_\rmi{i}^1 & \equiv
 \Tint{Q,R} \frac{(Q-P)^2}{Q^2R^2(Q-R)^2(R-P)^2}P_T(Q)
 \,,\displaybreak[0] \\
 \mathcal{I}_\rmi{i}^2 & \equiv
 \Tint{Q,R} \frac{P^2(Q-P)^2}{Q^2R^4(Q-R)^2(R-P)^2}P_T(Q)
 \,,\displaybreak[0] \\
 \mathcal{I}_\rmi{i}^3 & \equiv
 \Tint{Q,R} \frac{(Q-P)^4}{Q^2R^4(Q-R)^2(R-P)^2}P_T(R)
 \,,\displaybreak[0] \\
 \mathcal{I}_\rmi{i'} & \equiv
 \Tint{Q,R} \frac{4(Q\cdot P)^2}{Q^2R^2(Q-R)^2(R-P)^2}
 \,, \displaybreak[0]  \label{ii'}\\
 \mathcal{I}_\rmi{j}^0 & \equiv
 \Tint{Q,R} \frac{P^6}{Q^2R^2(Q-R)^2(Q-P)^2(R-P)^2}
 \,,\displaybreak[0] \\
 \mathcal{I}_\rmi{j}^1 & \equiv
 \Tint{Q,R} \frac{P^4}{Q^2R^2(Q-R)^2(Q-P)^2(R-P)^2}P_T(Q)
 \,,\displaybreak[0] \\
 \mathcal{I}_\rmi{j}^2 & \equiv
 \Tint{Q,R} \frac{P^4}{Q^2R^2(Q-R)^2(Q-P)^2(R-P)^2}P_T(Q-R)
 \,,\displaybreak[0] \\
 \mathcal{I}_\rmi{j}^3 & \equiv
 \Tint{Q,R} \frac{P^2}{Q^2R^2(Q-R)^2(Q-P)^2(R-P)^2}P_T(Q)^2
 \,,\displaybreak[0] \\
 \mathcal{I}_\rmi{j}^4 & \equiv
 \Tint{Q,R} \frac{P^2}{Q^2R^2(Q-R)^2(Q-P)^2(R-P)^2}P_T(Q-R)^2
 \,,\displaybreak[0] \\
 \mathcal{I}_\rmi{j}^5 & \equiv
 \Tint{Q,R} \frac{P^2}{Q^2R^2(Q-R)^2(Q-P)^2(R-P)^2}P_T(Q)P_T(R)
 \,,\displaybreak[0] \\
\mathcal{I}_\rmi{j}^6 & \equiv
 \Tint{Q,R} \frac{P^2}{Q^2R^2(Q-R)^2(Q-P)^2(R-P)^2}P_T(Q)P_T(Q-R)
 \,.
\end{align}
Following the notation of \cite{Zhu:2012be}, we have denoted here $P_T(Q) \equiv Q_\mu Q_\nu P^T_{\mu\nu}(P) = \mathbf{q}^2-(\mathbf{q\cdot \hat p})^2$, where $P^T_{\mu\nu}(P)$ is the usual three-dimensionally transverse projection operator defined with momentum $P$. The sum-integration measure used here is defined as
\ba
\Tint{Q} &\equiv& T \sum_{q_0} \int_\vec{q}, \quad \int_\vec{q}\, \equiv \,\int\! \frac{{\rm  d}^{D-1}\vec{q}}{(2\pi)^{D-1}} \,=\, \Lambda^{-2\epsilon} \left( \frac{{\rm e}^{\gamma_E}\bar\Lambda^2}{4\pi}\right)^{\epsilon}
\int\! \frac{{\rm  d}^{D-1}\vec{q}}{(2\pi)^{D-1}} \; ,
\ea
where $\Lambda$ and $\bar\Lambda$ stand for the renormalization scales in the MS and $\msbar$ schemes, respectively.

\subsection{Evaluation of the masters}

The methods required for the evaluation of the above sum-integrals were largely developed in the bulk calculation of \cite{Laine:2011xm} and later generalized to the shear channel integrals in \cite{Zhu:2012be}. As most of the above cases were already considered in these two references, we will only discuss the new cases in detail here. In subsection \ref{newcases}, we first evaluate the sum-integrals $\mathcal{I}_\rmi{h}^{4'}$, $\mathcal{I}_\rmi{h}^{5'}$, $\mathcal{I}_\rmi{h}^{6'}$, $\mathcal{I}_\rmi{h}^{7'}$ and $\mathcal{I}_\rmi{h}^{6*}$ that appear in the current computation due to our new way of treating the type (v) diagrams of fig.~1. After this, we will in subsection \ref{discre} revisit the evaluation of a few master integrals already encountered in \cite{Zhu:2012be}, explaining in detail a subtle problem we discovered in their original evaluation.

\subsubsection{New masters \label{newcases}}
%%%%%%%%%%%%%%%%%%%%%%%%%%%%%%%%%%

Following section 3 of \cite{Zhu:2012be}, we write our sum-integrals in the form 
\ba
 \rho^{ }_{\mathcal{I}^{n}_\rmii{x}}(\omega) \equiv 
 \int_{\vec{q,r}}f_{\mathcal{I}^{n}_\rmii{x}}\; , \label{fdef}
\ea
where the functions $f_{\mathcal{I}^{n}_\rmii{x}}$ read for the new masters 
\ba
 % Ih4'
 f_{\mathcal{I}^{4'}_\rmii{h}} &=&
 \frac{D(D-2)}{D^2-1}
  \frac{q^4}{\omega^4}
 f_{\mathcal{I}^{0}_\rmii{h}}\,,
 \\
 % Ih5'
 f_{\mathcal{I}^{5'}_\rmii{h}} &=&
 \frac{D(D-2)}{D^2-1}
 \frac{r^4}{\omega^4}
 f_{\mathcal{I}^{0}_\rmii{h}}\,,
 \\
 % Ih6'
 f_{\mathcal{I}^{6'}_\rmii{h}} &=&
 \left(\frac{D^2-2 D-2}{D^2-1}q^2r^2+\frac{2}{D^2-1}(\mathbf{q}\cdot\mathbf{r})^2\right) 
\frac{f_{\mathcal{I}^{0}_\rmii{h}}}{\omega^4} \; , \\
 % Ih6*
  f_{\mathcal{I}^{6*}_\rmii{h}} &=&
 \fr{q^2r^2}{\omega^4}
 f_{\mathcal{I}^{0}_\rmii{h}} \,,
 \\
 % Ih7'
 f_{\mathcal{I}^{7'}_\rmii{h}} &=&
  \left(\frac{D^2-2 D-2}{D^2-1}q^2(\mathbf{q-r})^2+\frac{2}{D^2-1}(\mathbf{q}\cdot(\mathbf{q-r}))^2\right) \frac{f_{\mathcal{I}^{0}_\rmii{h}}}{\omega^4}\, ,
\ea
and $f_{\mathcal{I}^{0}_\rmii{h}}$ is as defined in eq.~(A.37) of \cite{Zhu:2012be}. The evaluation of these integrals follows that of $\mathcal{I}_\rmi{h}^{4}$, $\mathcal{I}_\rmi{h}^{5}$, $\mathcal{I}_\rmi{h}^{6}$, and $\mathcal{I}_\rmi{h}^{7}$  step by step, except that no propagators need to be squared. We will thus simply list the corresponding results for the integrals below.

%%%%%%%%%%%%%%%%%%%%%%% SUBSUBSECTION %%%%%%%%%%%%%%%%%%%%%%%%%%%%%%%%%
\paragraph*{\texorpdfstring{$\rho^{ }_{\mathcal{I}^{4'}_\rmii{h}}(\omega)$}{}}
%%%%%%%%%%%%%% rho Ih4' %%%%%%%%%%%%%%%%%%%%%%%%%%%%%%%%%%%%%%%%%
The master $\mathcal{I}^{4'}_\rmii{h}$ is related to the non-differentiated version of $\mathcal{I}_\rmi{h}^{4}$, evaluated in eqs.~(B.55)--(B.57) of \cite{Zhu:2012be}. It is easy to verify that its so-called (fz,p) and (fz,e) parts (see \cite{Zhu:2012be} for definitions) read
\ba
  \rho^{(\rmi{fz,p})}_{\mathcal{I}^{4'}_\rmii{h}}(\omega)
 &=& 
 \frac{\omega^4\Lambda^{-4\epsilon}} {600(4\pi)^3}(1+2 n_{\frac{\omega}{2}}) 
\left(\frac{1}{\epsilon}+\ln \frac{\bmu^2}{m^2}+\ln \frac{\bmu^2}{(\omega-\fr{m^2}{\omega})^2}+\frac{81}{20}\right)
 \,, \la{Ih4p_fzp}\\
 \rho_{\mathcal{I}^{4'}_\rmii{h}}^{(\rmi{fz,e})}(\omega) &=&   
  \frac{8}{15} 
 \biggl[
  \frac{1}{2(4\pi)^3\omega} (1+n_{E_r}+n_r ) \\
  &\times&  \Bigg\{  
%  \nn && 
  \int_0^\infty \! {\rm d}q 
  \int_{E_{qr}^-}^{E_{qr}^+} \!\! {\rm d}E_{qr}  \, n_q \,
      \left(\frac{1}{\Delta_{00}} +\frac{1}{\Delta_{10}} -\frac{1}{\Delta_{01}} -\frac{1}{\Delta_{11}}\right) q^4
  \nn & + & 
  \int_{0}^\infty \! {\rm d}E_{qr}
  \int_{ |r - E_{qr}|}^{r + E_{qr}}
 \!\! {\rm d}q   \, n_{qr}
% \biggl[ 
    \left(\frac{1}{\Delta_{00}} +\frac{1}{\Delta_{01}} +\frac{1}{\Delta_{10}} +\frac{1}{\Delta_{11}}\right) q^4
       \Biggr\} 
  \biggl]_{r=\frac{\omega^2-m^2}{2\omega}}
  %%% 
  \,,  \la{Ih4_fz_e} \nonumber
\ea
where $\Delta_{ij}\equiv q+(-1)^{i}E_r+(-1)^{j}E_{qr}$, with $E_r=\sqrt{r^2+m^2}$ and $E_{qr}=|\vec{q}-\vec{r}|$. Finally, the (ps) part is obtained from eq.~(B.30) of this reference by simply inserting there the function
\ba
F_{\It{h}{4'}}(x,y,z)&=& \frac{8x^4}{15\omega^4}\, .
\ea

%%%%%%%%%%%%%%%%%%%%%%% SUBSUBSECTION %%%%%%%%%%%%%%%%%%%%%%%%%%%%%%%%%
\paragraph*{\texorpdfstring{$\rho^{ }_{\mathcal{I}^{5'}_\rmii{h}}(\omega)$}{}}
%%%%%%%%%%%%%% rho Ih5' %%%%%%%%%%%%%%%%%%%%%%%%%%%%%%%%%%%%%%%%%
Following the above strategy and using eqs.~(B.60)--(B.62) of \cite{Zhu:2012be}, we get
\ba
  \rho^{(\rmi{fz,p})}_{\mathcal{I}^{5'}_\rmii{h}}(\omega)
 &=& 
 \frac{\omega^4 \Lambda^{-4\epsilon}} {120(4\pi)^3}
 (1+2 n_{\frac{\omega}{2}})  \Bigg(
  \frac{1}{\epsilon}
 +\ln \frac{\bmu^2}{m^2}+\ln \frac{\bmu^2}{(\omega-\fr{m^2}{\omega})^2}
 +\frac{107}{30}
 \Biggr)
 \,, \la{Ih5_fzp} \\
 \rho_{\mathcal{I}^{5'}_\rmii{h}}^{(\rmi{fz,e})}(\omega) &=& 
 \frac{8}{15}
 \fr{1}{\omega^4}
 \left(\fr{\omega^2-m^2}{2\omega}\right)^4
 \rho_{\mathcal{I}^{0}_\rmii{h}}^{(\rmi{fz,e})}(\omega)
 \nn &=&  
  \frac{8}{15(4\pi)^3\omega} 
  \left(\fr{\omega^2-m^2}{2\omega}\right)^4
  (1+2n_{\fr{\omega}{2}} )
  \nn && \times
  \int_0^\infty \! {\rm d}q \, n_q \, 
  \ln\left|\fr{2q\omega-m^2}{2q\omega+m^2} \times \fr{2q+\omega}{2q-\omega}\right|
  \,,  \la{Ih5_fz_e}
\ea
as well as
\ba
F_{\It{h}{5'}}(x,y,z)&=& \frac{8y^4}{15\omega^4}\, .
\ea
%%%%%%%%%%%%%%%%%%%%%%% SUBSUBSECTION %%%%%%%%%%%%%%%%%%%%%%%%%%%%%%%%%
\paragraph*{\texorpdfstring{$\rho^{ }_{\mathcal{I}^{6'}_\rmii{h}}(\omega)$}{}}
%%%%%%%%%%%%%% rho Ih6' %%%%%%%%%%%%%%%%%%%%%%%%%%%%%%%%%%%%%%%%%
In this case, the (fz,p) part can be directly read off from eq.~(B.65) of \cite{Zhu:2012be} as
\ba
  \rho^{(\rmi{fz,p})}_{\mathcal{I}^{6'}_\rmii{h}}(\omega,m)
 &=& 
 \frac{\omega^4\Lambda^{-4\epsilon}} {360(4\pi)^3}
 (1+2 n_{\frac{\omega}{2}}) 
 \left(\frac{1}{\epsilon}+\ln \frac{\bmu^2}{m^2}+\ln \frac{\bmu^2}{(\omega-\fr{m^2}{\omega})^2}+\frac{56}{15}\right)
 \,. 
\ea
For the other parts, we on the other hand use the relation
\ba
 % Ih6'
 \rho^{ }_{\mathcal{I}^{6'}_\rmii{h}}(\omega) &=&
 \frac{1}{2D(D-2)}\rho^{ }_{\mathcal{I}^{4'}_\rmii{h}}(\omega)
 +\frac{1}{2D(D-2)}\rho^{ }_{\mathcal{I}^{5'}_\rmii{h}}(\omega)
 \nn &+&
\fr{1}{\omega^2}\bigg(\rho^{ }_{\mathcal{I}^{6(1)}_\rmii{h}}(\omega,m)+\rho^{ }_{\mathcal{I}^{6(2)}_\rmii{h}}(\omega,m)+\rho^{ }_{\mathcal{I}^{6(3)}_\rmii{h}}(\omega,m)+\rho^{ }_{\mathcal{I}^{6(4)}_\rmii{h}}(\omega,m)\bigg)\,,
\ea
and the fact that the (fz,e) and (ps) parts of the $\rho^{ }_{\mathcal{I}^{6(n)}_\rmii{h}}(\omega,m)$ are listed in eqs.~(B.71)--(B.79) of \cite{Zhu:2012be}.

%%%%%%%%%%%%%%%%%%%%%%% SUBSUBSECTION %%%%%%%%%%%%%%%%%%%%%%%%%%%%%%%%%
\paragraph*{\texorpdfstring{$\rho^{ }_{\mathcal{I}^{6*}_\rmii{h}}(\omega)$}{}}
%%%%%%%%%%%%%% rho Ih6* %%%%%%%%%%%%%%%%%%%%%%%%%%%%%%%%%%%%%%%%%
The new master $\rho^{ }_{\mathcal{I}^{6*}_\rmii{h}}(\omega)$ is related to $\rho^{ }_{\mathcal{I}^{6(1)}_\rmii{h}}(\omega,m)$ of \cite{Zhu:2012be}, via 
\be
\rho_{\mathcal{I}^{6*}_\rmii{h}}(\omega)\equiv\fr{D^2-1}{(D^2-2D-1)}\fr{\rho_{\mathcal{I}^{6(1)}_\rmii{h}}(\omega,m)}{\omega^2}\, .
\ee
Using this result, we immediately obtain from \cite{Zhu:2012be}
\ba
  \rho^{(\rmi{fz,p})}_{\mathcal{I}^{6*}_\rmii{h}}(\omega)
 &=& 
 \frac{\omega^4\Lambda^{-4\epsilon}} {192(4\pi)^3}
 (1+2 n_{\frac{\omega}{2}}) 
 \left(\frac{1}{\epsilon}+2\ln \frac{\bmu^2}{\omega^2}+\frac{57}{10}\right)
 \,,
\ea
as well as
\be
\rho^{(\rmi{fz,e})}_{\mathcal{I}^{6*}_\rmii{h}}(\omega)\equiv\fr{D^2-1}{(D^2-2D-1)}\fr{\rho^{(\rmi{fz,e})}_{\mathcal{I}^{6(1)}_\rmii{h}}(\omega,m)}{\omega^2}\,,\hspace{1cm}
F_{\It{h}{6*}}(x,y,z)= \frac{x^2y^2}{\omega^4}\, .
\ee

%%%%%%%%%%%%%%%%%%%%%%% SUBSUBSECTION %%%%%%%%%%%%%%%%%%%%%%%%%%%%%%%%%
\paragraph*{\texorpdfstring{$\rho^{ }_{\mathcal{I}^{7'}_\rmii{h}}(\omega)$}{}}
%%%%%%%%%%%%%% rho Ih7' %%%%%%%%%%%%%%%%%%%%%%%%%%%%%%%%%%%%%%%%%
For the last of our new masters, the (fz,p) part can be read from eq.~(B.82) of \cite{Zhu:2012be}
\ba
  \rho^{(\rmi{fz,p})}_{\mathcal{I}^{7'}_\rmii{h}}(\omega)
 &=& 
 \frac{\omega^4\Lambda^{-4\epsilon}} {3600(4\pi)^3}(1+2 n_{\frac{\omega}{2}}) 
 \left(\frac{1}{\epsilon}+\ln \frac{\bmu^2}{m^2}+\ln \frac{\bmu^2}{(\omega-\fr{m^2}{\omega})^2}+\fr{94}{3}\right)
 \,, \nonumber
\ea
while (fz,e) and (ps) parts are available using the identity
\ba
 \rho^{ }_{\mathcal{I}^{7'}_\rmii{h}}(\omega) 
 &= &\rho^{ }_{\mathcal{I}^{6'}_\rmii{h}}(\omega)
 - \frac{D (D-2)}{D^2-2D-1}\fr{\rho^{ }_{\mathcal{I}^{6(1)}_\rmii{h}}(\omega,m)}{\omega^2}
 - D (D-2)\fr{\rho^{ }_{\mathcal{I}^{6(3)}_\rmii{h}}(\omega,m)}{\omega^2}
 \,. \la{Ih7_exp}
\ea

\subsubsection{Discrepancy with previous results \label{discre}}
%%%%%%%%%%%%%%%%%%%%%%%%%%%%%%%%%%
Next, we move on to discuss the tension between our new results and those of ref.~\cite{Zhu:2012be}. A lengthy independent calculation of the contributions of the different graphs of fig.~1 to the shear spectral function revealed a numerical discrepancy in the results of the type (v) diagrams. Upon closer inspection, the source of the problem was isolated to those masters of \cite{Zhu:2012be} containing a squared propagator, which indeed only appear in type (v) graphs. In our earlier work, our method of dealing with them consisted of introducing an auxiliary mass parameter $m^2$ in the propagator in question, and then making use of the simple identity 
\be
\fr{1}{R^4}=-\lim_{m\to 0} 
 \Bigg\{
  \frac{{\rm d}}{{\rm d}m^2}\fr{1}{R^2+m^2}
   \Biggr\}\, .
\ee
As discussed in appendix B of \cite{Zhu:2012be}, in the $m\to 0$ limit one encounters IR divergences in the one- and two-dimensional integrals originating from the (fz,e) and (ps) parts of these masters. Our strategy with them was to separate out the divergent terms and treat them analytically, while the finite remainder was computed numerically after setting $m$ to 0. 

Unfortunately, it turns out that the procedure applied in \cite{Zhu:2012be} missed a set of finite contributions that would have been correctly accounted for had we managed to set $m$ to zero only after performing all the integrations. To see this in detail, consider the one-dimensional part of $\rho^{ }_{\mathcal{I}^{3}_\rmii{h}}(\omega)$, which in \cite{Zhu:2012be} was written in the form
\ba
\fr{(4\pi )^3\rho^{(\rmi{1d}) }_{\mathcal{I}^{3}_\rmii{h}}(\omega)} { 1+2n_{\fr{\omega}{2}}}
&=& \Bigg\{\fr{(4\pi )^3\rho^{(\rmi{1d}) }_{\mathcal{I}^{3}_\rmii{h}}(\omega)} { 1+2n_{\fr{\omega}{2}}}-\int_{ \frac{\omega}{2} }^{\infty}
 \! {\rm d}q \, 
 n_{\fr{\omega}{2}} \frac{\omega }{2q}
 \nn&\times&\Bigg[
 \frac{-m^4+\omega ^4}{24 \left(q+\frac{m^2}{2 \omega }-\frac{\omega }{2}\right)}+\frac{m^6-2 m^4 \omega ^2+m^2 \omega ^4}{48 \left(q+\frac{m^2}{2 \omega }-\frac{\omega }{2}\right)^2 \omega }
 \Bigg]
 \Bigg\}
 \nn &+&
\int_{ \frac{\omega}{2} }^{\infty}
 \! {\rm d}q \;n_{\fr{\omega}{2}} \frac{\omega }{2q}
 \Bigg[
 \frac{-m^4+\omega ^4}{24 \left(q+\frac{m^2}{2 \omega }-\frac{\omega }{2}\right)}+\frac{m^6-2 m^4 \omega ^2+m^2 \omega ^4}{48 \left(q+\frac{m^2}{2 \omega }-\frac{\omega }{2}\right)^2 \omega }
 \Bigg]
 \,. \nonumber
\ea
Of the terms here, we evaluated the integral in the last row analytically, while the terms on the first two rows were treated numerically after first setting $m\to 0$ inside the integrand. The problem with this are the IR divergent terms in the subtracted part that are thrown out due to being explicitly proportional to $m^2$. For some of them, the actions of performing the integrations and taking the $m\to 0$ limit do not commute, as can be verified e.g.~from the analytic result
\be
\lim_{m\to 0} 
 \Bigg\{-\int_{ \frac{\omega}{2} }^{\infty}
 \! {\rm d}q \;n_{\fr{\omega}{2}}\frac{\omega }{2q}\frac{m^6-2 m^4 \omega ^2+m^2 \omega ^4}{48 \left(q+\frac{m^2}{2 \omega }-\frac{\omega }{2}\right)^2 \omega } \Bigg\}
 =
 -\frac{\omega ^4}{24}n_{\frac{\omega }{2}}\,.
\ee
Similar cases also appear in the two-dimensional integrals encountered in the calculation.

A careful treatment of all the master integrals with squared propagators reveals that the results of \cite{Zhu:2012be} must be supplemented with the extra contributions
\ba
\fr{(4\pi )^3\rho^{(\rmi{1d,extra}) }_{\mathcal{I}^{3}_\rmii{h}}(\omega)} { 1+2n_{\fr{\omega}{2}}} &=& -\frac{\omega ^4}{24}n_{\frac{\omega }{2}}\,,
\\
\fr{(4\pi )^3\rho^{(\rmi{2d,extra}) }_{\mathcal{I}^{3}_\rmii{h}}(\omega)} { 1+2n_{\fr{\omega}{2}}} &=& \frac{\omega ^4}{24}n_{\frac{\omega }{2}}\,,
\\
\fr{(4\pi )^3\rho^{(\rmi{2d,extra}) }_{\mathcal{I}^{4}_\rmii{h}}(\omega)} { 1+2n_{\fr{\omega}{2}}} &=&-\frac{16 \pi ^4T^4}{225}-\frac{2 \pi ^2 \omega ^2T^2}{15}+\frac{\omega ^4}{120}+\frac{16}{5}T^4 \text{Li}_4\left[e^{\omega /(2T)}\right] \,, 
\\
\fr{(4\pi )^3\rho^{(\rmi{1d,extra}) }_{\mathcal{I}^{5}_\rmii{h}}(\omega)} { 1+2n_{\fr{\omega}{2}}} &=& \frac{\omega ^4}{120}n_{\frac{\omega }{2}}\,,
\\
\fr{(4\pi )^3\rho^{(\rmi{2d,extra}) }_{\mathcal{I}^{5}_\rmii{h}}(\omega)} { 1+2n_{\fr{\omega}{2}}} &=&-\frac{\omega ^4}{120}n_{\frac{\omega }{2}} \,, 
\\
\fr{(4\pi )^3\rho^{(\rmi{2d,extra}) }_{\mathcal{I}^{6}_\rmii{h}}(\omega)} { 1+2n_{\fr{\omega}{2}}} &=&-\frac{1}{45} \pi ^2 \omega ^2T^2+\frac{\omega ^4}{120}+\frac{1}{15} \omega ^2 T^2\text{Li}_2\left[e^{\omega /(2T)}\right] \,, 
\\
\fr{(4\pi )^3\rho^{(\rmi{2d,extra}) }_{\mathcal{I}^{7}_\rmii{h}}(\omega)} { 1+2n_{\fr{\omega}{2}}} &=&-\frac{16 \pi ^4T^4}{225}-\frac{\pi ^2 \omega ^2T^2}{45}+\frac{1}{15} \omega ^2T^2 \text{Li}_2\left[e^{\omega /(2T)}\right]
\nn&&-\frac{4}{5} \omega T^3  \text{Li}_3\left[e^{\omega /(2T)}\right]+\frac{16}{5} T^4 \text{Li}_4\left[e^{\omega /(2T)}\right] \,, 
\\
\fr{(4\pi )^3\rho^{(\rmi{2d,extra}) }_{\mathcal{I}^{2}_\rmii{i}}(\omega)} { 1+2n_{\fr{\omega}{2}}} &=&-\frac{2}{9} \pi ^2 \omega ^2T^2-\frac{1}{3} \omega ^2 T^2\text{Li}_2\left[e^{\omega /(2T)}\right]
\nn&&+2 \omega T^3  \text{Li}_3\left[e^{\omega /(2T)}\right]\,, \\
\fr{(4\pi )^3\rho^{(\rmi{2d,extra}) }_{\mathcal{I}^{3}_\rmii{i}}(\omega)} { 1+2n_{\fr{\omega}{2}}} &=&\frac{1}{3} \omega^2T^2  \text{Li}_2\left[e^{-\omega /(2T)}\right]
\,,
\ea
where Li stands for the polylogarithmic function. Once all of these terms are taken into account, one can straightforwardly verify that the discrepancy between the results of \cite{Zhu:2012be} and our new computation has completely vanished.

\section{Master integrals in the naive HTL calculation}\label{htl1}
%%%%%%%%%%%%%%%%%%%%%%%%%%%%%%%%%%
The master integrals appearing in the naive HTL calculation of section \ref{htl_naive} can be evaluated using techniques familiar from \cite{Laine:2011xm}. First carrying out the Matsubara sums with standard methods and then taking the imaginary parts of the results, we obtain the simple intermediate results 
\ba
&&\hspace{-1.5cm} \im \biggl\{
 \Tint{Q} \frac{q^m}{[Q^2+\lambda^2] [(Q-P)^2+\lambda^2]}
 \biggr\}_{P \to (-i [\omega + i 0^+],\vec{0})}
 \nn
 & = &  
\int_{\vec{q}} \frac{\pi q^m}{4 E_q^2}
 [\delta(\omega - 2 E_q) - \delta(\omega + 2 E_q)]
 \bigl( 1+2 n^{ }_{E_q} \bigr)
 \nn & = &   
   \left[\frac{ \pi ^{(D-1)/2}}{(2\pi )^{D-1}\Gamma\left(\fr{D-1}{2}\right)}\fr{\pi q^{D-3+m}}{4 E_q}
  \bigl( 1 + 2 n_{\frac{\omega}{2}}\bigr)
  \,\theta(\omega - 2\lambda) + \rmO(\epsilon)\right]_{q=\left[(\fr{\omega}{2})^2-\lambda^2\right]^{\fr12}}
 \,, \la{Im1}\\
%%%%%%%%%%%%%
 & & \hspace*{-1.5cm} \im \biggl\{
 \Tint{Q}\int_z \frac{q^m }{[Q^2+\lambda^2](Q-P)^2}   \frac{q z}{iq_n + q z} 
 \biggr\}_{P \to (-i [\omega + i 0^+],\vec{0})}
 \nn & = & 
 \frac{1}{8\pi}
 \biggl\{ 
  \int_{\frac{\omega}{2}}^{\omega} 
  \! {\rm d}q \, \frac{q^m(\omega - q)}{2\omega q - \omega^2}
  \bigl(1 + 2 n_{\frac{\omega}{2}} \bigr)
  \frac{n_q (1 + n_{\omega - q})}{n^2_{q-\frac{\omega}{2}} }
 \nn & & \qquad - \, 
  \int_{\omega}^{\infty} 
  \! {\rm d}q \, \frac{q^m(\omega - q)}{2\omega q - \omega^2}
   \bigl(1 + 2 n_{\frac{\omega}{2}} \bigr)
  \frac{n_q (1 + n_{q-\omega})}{n^2_{\frac{\omega}{2}} }
 \nn & & \qquad + \, 
  \int_{\frac{\omega}{2}}^{\omega} 
  \! {\rm d}q \, \frac{q^m(\omega - q)}{\lambda^2 + 2\omega q - \omega^2}
  \bigl( 1 +  2 n_{\frac{\omega}{2}} \bigr)
 \nn & & \qquad + \, 
  \left. \frac{q^mE_q}{q + E_q}
  \biggl( \frac{q}{E_q} + \fr12 \ln\frac{E_q - q}{E_q + q}  \biggr)
  \bigl( 1+ n_q + n^{ }_{E_q} \bigr)
  \right|_{q = \frac{\omega^2 - \lambda^2}{2\omega}}
 \biggr\} 
 \,,   \la{Im2}\\
%%%%%%%%%%%%%%%
 & & \hspace*{-1.5cm} \im \biggl\{
 \Tint{Q}\int_z \frac{q^m }{(Q-P)^2}   \frac{q z}{iq_n + q z} 
 \biggr\}_{P \to (-i [\omega + i 0^+],\vec{0})}
 \nn & = & 
 -\frac{1}{8\pi}
  \int_{\fr{\omega}{2}}^{\infty} 
  \! {\rm d}q \, q^m(\omega - q)
   \bigl(1 + 2 n_{\frac{\omega}{2}} \bigr)
  \frac{n_q (1 + n_{q-\omega})}{n^2_{\frac{\omega}{2}} }
 \,, \la{Im3} 
 \ea
where $m$ is a non-negative integer, $\lambda$ a regulatory mass parameter, and $E_q\equiv\sqrt{q^2+\lambda^2}$. Taking then a derivative with respect to $\lambda$ (and making sure that no issues such as those explained in the previous section arise), we further obtain the identities
\ba
  \Tint{Q} \frac{q^m}{Q^4(Q-P)^2}
  & = & 
  - \fr12\lim_{\lambda\to 0} \frac{{\rm d}}{{\rm d}\lambda^2}
 \Tint{Q}  
  \frac{q^m}{[Q^2+\lambda^2][(Q-P)^2+\lambda^2]}\,, \\
  %%%%
  \Tint{Q}\int_z \frac{q^m}{Q^4(Q-P)^2}  \frac{q z}{iq_n + q z} 
  & = & 
  - \lim_{\lambda\to 0} \frac{{\rm d}}{{\rm d}\lambda^2}
 \Tint{Q} \int_z  
  \frac{q^m}{[Q^2+\lambda^2](Q-P)^2}  \frac{q z}{iq_n + q z} \,,
\ea
which leads us to the following results for the master integrals appearing in eq.~\nr{Geta_naive}:
\ba
%%%
&&\hspace{-2cm} \im \biggl\{ \Tint{Q}\frac{\Pi^\text{HTL} _T(Q)q^2p_n{}^2}{Q^4(Q-P)^2} \biggr\}_{P \to (-i [\omega + i 0^+],\vec{0})}=  -\frac{\mE^2(1+2n_{\frac{\omega }{2}})}{16\pi}  \biggl\{\frac{ \omega ^2 }{8 } 
\nn && + \int_{\fr{\omega}{2}}^{\omega} {\rm d}q \frac{ \omega (q-\omega )  n_q (1+n_{\omega-q })}{(2 q-\omega ) n_{q-\frac{\omega }{2}}^2}  +  \int_{\omega}^{\infty} {\rm d}q \frac{\omega  (\omega-q ) n_q (1+n_{q-\omega })}{(2 q-\omega ) n_{\frac{\omega }{2}}^2}\biggr\} \,,\la{I1}\\
%%%
&&\hspace{-2cm} \im \biggl\{ \Tint{Q}\frac{\Pi^\text{HTL} _T(Q)q^4}{Q^4(Q-P)^2} \biggr\}_{P \to (-i [\omega + i 0^+],\vec{0})}= \frac{\mE^2(1+2n_{\frac{\omega }{2}})}{16\pi}  \biggl\{\frac{ \omega ^2 }{48 } 
\nn && + \int_{\fr{\omega}{2}}^{\omega} {\rm d}q \frac{ q^2 (q-\omega )  n_q (1+n_{\omega-q })}{ \omega (2 q-\omega ) n_{q-\frac{\omega }{2}}^2}  +  \int_{\omega}^{\infty} {\rm d}q \frac{ q^2 (\omega-q ) n_q (1+n_{q-\omega })}{\omega (2 q-\omega ) n_{\frac{\omega }{2}}^2}\biggr\} \,,\la{I2}\\
%%%
%&&\hspace{-2cm} \im \biggl\{ \Tint{Q} \frac{\Pi^\text{HTL} _T(Q)p_n{}^2}{Q^2(Q-P)^2} \biggr\}_{P \to (-i [\omega + i 0^+],\vec{0})}=  -\frac{\mE^2(1+2n_{\frac{\omega }{2}})}{16\pi}  \biggl\{\frac{ \omega ^2 }{2 }
% \nn && +  \int_{\frac{\omega }{2}}^{\infty} {\rm d}q \frac{ \omega^2 (\omega-q ) n_q (1+n_{q-\omega })}{q^2 n_{\frac{\omega }{2}}^2}\biggr\} \,, \la{I3}\\
%%%
&&\hspace{-2cm} \im \biggl\{ \Tint{Q}\frac{\Pi^\text{HTL} _T(Q)q^2}{Q^2(Q-P)^2} \biggr\}_{P \to (-i [\omega + i 0^+],\vec{0})}=  \frac{\mE^2(1+2n_{\frac{\omega }{2}})}{16\pi}  \biggl\{\frac{ \omega ^2 }{8 } 
 \nn && +  \int_{\frac{\omega }{2}}^{\infty} {\rm d}q \frac{ (\omega-q ) n_q (1+n_{q-\omega })}{n_{\frac{\omega }{2}}^2}\biggr\} \,, \la{I4}\\
%%%
%&&\hspace{-2cm} \im \biggl\{ \Tint{Q} \frac{p_n{}^2}{Q^2(Q-P)^2} \biggr\}_{P \to (-i [\omega + i 0^+],\vec{0})}= -\frac{\omega ^2}{16 \pi } (1+2n_{\frac{\omega }{2}}) \,, \\
%%%
&&\hspace{-2cm} \im \biggl\{ \Tint{Q}\frac{q^2}{Q^2(Q-P)^2} \biggr\}_{P \to (-i [\omega + i 0^+],\vec{0})}= \frac{\omega ^2}{64 \pi } (1+2n_{\frac{\omega }{2}}) \,, \\
%%%
&&\hspace{-2cm} \im \biggl\{ \Tint{Q} \frac{q^2p_n{}^2}{Q^2(Q-P)^2} \biggr\}_{P \to (-i [\omega + i 0^+],\vec{0})}= -\frac{\omega ^4}{64 \pi } (1+2n_{\frac{\omega }{2}}) \,,\\
%%%
&&\hspace{-2cm} \im \biggl\{ \Tint{Q}\frac{q^4}{Q^2(Q-P)^2} \biggr\}_{P \to (-i [\omega + i 0^+],\vec{0})}= \frac{\omega ^4}{256 \pi } (1+2n_{\frac{\omega }{2}})\,.
\ea

As the HTL resummation is intended to be carried out in the region $\omega\ll T$, we can simplify the above results by carrying out an expansion in positive powers of $\omega$. For the integrals with the range $\int_{\fr{\omega}{2}}^{\omega}{\rm d}q$, this is easily accomplished by taking the classical limit $(1+)n_q\to T/q$, which leads to the simple results
\ba
\int_{\fr{\omega}{2}}^{\omega} {\rm d}q \frac{ \omega (q-\omega )  n_q (1+n_{\omega-q })}{(2 q-\omega ) n_{q-\frac{\omega }{2}}^2}
&=&-\int_{\fr{\omega}{2}}^{\omega} {\rm d}q \frac{ \omega (2q-\omega )}{4q}+\mathcal{O}(\omega^4)
\nn
&=&\fr{\omega^2}{4}(\ln 2-1)+\mathcal{O}(\omega^4)
\,, \\
\int_{\fr{\omega}{2}}^{\omega} {\rm d}q \frac{ q^2 (q-\omega )  n_q (1+n_{\omega-q })}{ \omega (2 q-\omega ) n_{q-\frac{\omega }{2}}^2}
&=&-\fr{5\omega^2}{96}+\mathcal{O}(\omega^4)\,.
\ea
For the $q$ integrals extending to infinity, we on the other hand split the integration range to two parts,  $\int_{}^{\infty}{\rm d}q= \int_{}^{\Lambda}{\rm d}q+ \int_{\Lambda}^{\infty}{\rm d}q$, where $\Lambda$ is assumed to satisfy $\omega \ll \Lambda \ll T$. On the first of these ranges, we can again apply the `classical' replacement of $(1+)n_q\to T/q$, while on the latter one, we expand the integrand in powers of $\omega$. Taking the last term of eq.~\nr{I2} as an example, we obtain for the two integration regions
\ba
&&\hspace{-1.0cm}\int_{\omega}^{\Lambda} {\rm d}q \frac{ q^2 (\omega-q ) n_q (1+n_{q-\omega })}{\omega (2 q-\omega ) n_{\frac{\omega }{2}}^2} 
=-\int_{\omega}^{\Lambda} {\rm d}q \frac{q\omega}{4(2q-\omega)}+\mathcal{O}(\omega^3)
\nn&=&\fr{\omega^2}{16}(-\fr{2\Lambda}{\omega}+2+\omega\ln\fr{\omega}{2\Lambda}) +\mathcal{O}(\omega^3)\,,\\
&&\hspace{-1.0cm}\int_{\Lambda}^{\infty} {\rm d}q \frac{ q^2 (\omega-q ) n_q (1+n_{q-\omega })}{\omega (2 q-\omega ) n_{\frac{\omega }{2}}^2}
\nn&=&-\int_{\Lambda}^{\infty} {\rm d}q \biggl\{\fr{q^2e^{q/T}}{8(e^{q/T}-1)^2}\fr{\omega}{T^2}+\fr{q\,e^{q/T}[q(1+e^{q/T})+T(1-e^{q/T})]}{16(e^{q/T}-1)^3}\fr{\omega^2}{T^3}\biggr\} +\mathcal{O}(\omega^3)
\nn&=&-\fr{\omega\pi^2 T}{24}+\fr{\omega^2}{16}(\ln\fr{\Lambda}{T}-2)+\fr{\Lambda\omega}{8}+\mathcal{O}(\omega^3)\,,
\ea
the sum of which is clearly independent of $\Lambda$. This quickly leads to the results
\ba
\int_{\omega}^{\infty} {\rm d}q \frac{\omega  (\omega-q ) n_q (1+n_{q-\omega })}{(2 q-\omega ) n_{\frac{\omega }{2}}^2}
&=&-\fr{\omega^2\ln 2}{4}+\mathcal{O}(\omega^3)\,, \\
%%%
\int_{\omega}^{\infty} {\rm d}q \frac{ q^2 (\omega-q ) n_q (1+n_{q-\omega })}{\omega (2 q-\omega ) n_{\frac{\omega }{2}}^2}
&=&-\fr{\omega\pi^2 T}{24}+\fr{\omega^2}{16}\ln\fr{\omega}{2T}+\mathcal{O}(\omega^3)\,, \\
\int_{\frac{\omega }{2}}^{\infty} {\rm d}q \frac{ (\omega-q ) n_q (1+n_{q-\omega })}{n_{\frac{\omega }{2}}^2}
&=&-\fr{\omega^2}{4}(1-\ln\fr{\omega}{2T})+\mathcal{O}(\omega^3)\,,
\ea
which ultimately provides us with the final result for the naive HTL contribution to the shear spectral function, eq.~\nr{result1x}.
 
%%%%%%%%%%%%%%%%%%%%%%%%%%%%%%%%%%%%%%%%%%%%%%
\section{Resummed HTL calculation}\la{htl2}
%%%%%%%%%%%%%%%%%%%%%%%%%%%%%%%%%%%%%%%%%%%%%%

In the evaluation of the resummed HTL contribution to the shear spectral function, one encounters the functions (cf.~appendix C of \cite{Laine:2011xm})
\ba
&& \rho^{ }_T (q^0,q) \equiv
 \im \biggl\{ \frac{1}{\Delta^{ }_T(q_n,q)} \biggr\}_{q_n\to -i[q^0+i0^+]}
  =  
 \left\{
   \begin{array}{ll} 
      \displaystyle\frac{\Gamma^{ }_T(\eta)}
      {\Sigma^2_T(\mathcal{Q})+\Gamma^2_T(\eta)} \,,  &
      |\eta| < 1 \,, \\[3mm]
      \displaystyle
      \pi \mathop{\mbox{sign}} (\eta)
      \, \delta(\Sigma^{ }_T(\mathcal{Q})) \,, & 
      |\eta| > 1 \,, 
   \end{array} 
 \right.  \la{rho_T}
 \,, \\ 
&& \rho^{ }_E (q^0,q) \equiv
 \im \biggl\{ \frac{1}{\Delta^{ }_E(q_n,q)} \biggr\}_{q_n\to -i[q^0+i0^+]}
  =  
 \left\{
   \begin{array}{ll} 
      \displaystyle\fr{1}{\eta^2-1}\frac{\Gamma^{ }_E(\eta)}
      {\Sigma^2_E(\mathcal{Q})+\Gamma^2_E(\eta)} \,,  &
      |\eta| < 1 \,, \\[3mm]
      \displaystyle
      \fr{\pi \mathop{\mbox{sign}} (\eta)
      \, \delta(\Sigma^{ }_E(\mathcal{Q}))} {\eta^2-1} \,, & 
      |\eta| > 1 \,, 
   \end{array} 
 \right.  \la{rho_E} \\
&& \hat\rho^{ }_T (q^0,q) \equiv
 \im \biggl\{ \frac{\Pi^\text{HTL}_T(q_n,q)}{\Delta^{ }_T(q_n,q)} 
   \biggr\}_{q_n\to -i[q^0+i0^+]}
 =  
% \left\{
%   \begin{array}{ll} 
%      \displaystyle\frac{\mathcal{Q}^2\, \Gamma^{ }_T(\eta)}
%      {\Sigma^2_T(\mathcal{Q})+\Gamma^2_T(\eta)} \,,  &
%      |\eta| < 1 \,, \\[3mm]
%      \displaystyle
%      \pi \mathop{\mbox{sign}} (\eta)\, \mathcal{Q}^2
%      \, \delta(\Sigma^{ }_T(\mathcal{Q})) \,, & 
%      |\eta| > 1 \,, 
%   \end{array} \la{rho_T1}
% \right. \\ 
 \mathcal{Q}^2 \rho^{ }_T (q^0,q) \,, \\
&& \hat\rho^{ }_E (q^0,q) \equiv
 \im \biggl\{ \frac{\Pi^\text{HTL}_E(q_n,q)}{\Delta^{ }_E(q_n,q)}
   \biggr\}_{q_n\to -i[q^0+i0^+]}
    =  
% \left\{
%   \begin{array}{ll} 
%      \displaystyle\frac{q^2\, \Gamma^{ }_E(\eta)}
%      {\Sigma^2_E(\mathcal{Q})+\Gamma^2_E(\eta)} \,,  &
%      |\eta| < 1 \,, \\[3mm]
%      \displaystyle
%      \pi \mathop{\mbox{sign}} (\eta)\, q^2
%      \, \delta(\Sigma^{ }_E(\mathcal{Q})) \,, & 
%      |\eta| > 1 \,, 
%   \end{array} 
% \right. \la{rho_E1}%\quad
 \mathcal{Q}^2 \rho^{ }_E (q^0,q) \,,
\ea
where we have introduced the notation
\be
 \mathcal{Q} \equiv (q^0,\vec{q}) \,, \quad
 \mathcal{Q}^2 \equiv (q^0)^2 - q^2 \,, \quad
 \eta \equiv \frac{q^0}{q}
 \,, 
\ee
as well as denoted
\ba
 \Sigma^{ }_T(\mathcal{Q}) & \equiv & 
 -\mathcal{Q}^2 + \frac{\mE^2}{2}\biggl[
  \eta^2 + \frac{\eta(1-\eta^2)}{2}
  \ln\left| \frac{1+\eta}{1-\eta} \right| \biggr] 
 \,, \la{SigT} \\
 \Gamma^{ }_T(\eta) & \equiv & \frac{\pi \mE^2 \eta (1-\eta^2)}{4}
 \,, \la{GamT} \\
%%%%%%
 \Sigma^{ }_E(\mathcal{Q}) & \equiv & 
  q^2 + \mE^2 \biggl[ 1 - \frac{\eta}{2}
  \ln\left| \frac{1+\eta}{1-\eta} \right| \biggr] 
 \,, \la{SigE} \\
 \Gamma^{ }_E(\eta) & \equiv & \frac{\pi \mE^2 \eta}{2}
 \,. \la{GamE} 
\ea
Upon inserting these expressions into the integral of eq.~(\ref{rho_HTL_full}), the calculation reduces to one almost identical to that performed in Appendix C of \cite{Laine:2011xm}. We thus refrain from providing further details of this rather straightforward exercise, and simply display its numerical outcome in section \ref{HTLresults}.

\end{appendix}

\end{document}